\documentclass[twocolumn,superscriptaddress,
amsmath,amssymb,
longbibliography,reprint]{revtex4-2}

\usepackage{datetime2}

\usepackage{amssymb,amsmath}
\DeclareMathAlphabet\mathbfcal{OMS}{cmsy}{b}{n}
\usepackage{cmap}
\usepackage{graphicx} 
\usepackage{bm}% bold math
\usepackage{color}
\usepackage{blindtext}
\usepackage{hyperref}
\usepackage{listings}
\usepackage{booktabs}
\usepackage{multirow}
\usepackage{amsmath}
\usepackage{mathrsfs}
\usepackage{braket}
\usepackage{mathtools}

\usepackage{dsfont}
\usepackage{tabularx}

\newcommand{\red}[1]{{\color{black}{#1}}}
\newcommand{\blue}[1]{{\color{black}{#1}}}

\begin{document}

%\title{\large{Geometric and Dynamic Parts of the Generalized Wigner-Smith Operator}}

%\title{\large{Geometric and Dynamic Parts of the Generalized Wigner-Smith Operators \\ and Conjugate Variable Shifts in Wave Scattering}}

%\title{\large{Generalized Wigner-Smith Operators for Dynamic and Geometric \\ Phenomena in Wave Scattering}}

\title{{Dynamic and Geometric Shifts in Wave Scattering}}

%\title{\large{Geometric and Dynamic parts of Conjugate Variable Shifts in Wave Scattering}}

\author{{Konstantin Y. Bliokh}}
\affiliation{Donostia International Physics Center (DIPC), Donostia-San Sebasti\'{a}n 20018, Spain}
\affiliation{IKERBASQUE, Basque Foundation for Science, Bilbao 48009, Spain}
\affiliation{Centre of Excellence ENSEMBLE3 Sp.~z o.o., 01-919 Warsaw, Poland}

\author{{Zeyu Kuang}}
\affiliation{Institute for Theoretical Physics, TU Wien, Wiedner Hauptstraße 8-10/136, 1040 Vienna, Austria}

\author{{Stefan Rotter}}
\affiliation{Institute for Theoretical Physics, TU Wien, Wiedner Hauptstraße 8-10/136, 1040 Vienna, Austria}

%\vspace{1cm}
%\date{\today}

\begin{abstract}
Since Berry's pioneering 1984 work, the separation of \textit{geometric} and \textit{dynamic} contributions in the {\it phase} of an evolving wave has become fundamental in physics, underpinning diverse phenomena in quantum mechanics, optics, and condensed matter. Here we extend this {geometric-dynamic} decomposition from the wave-evolution phase to a distinct class of {\it wave scattering} problems, where observables (such as frequency, momentum, or position) experience {\it shifts in their expectation values} between the input and output wave states. We describe this class of problems using a unitary scattering matrix and the associated \textit{generalized Wigner-Smith operator} (GWSO), which involves gradients of the scattering matrix with respect to conjugate variables (time, position, or momentum, respectively). We show that both the GWSO and the resulting expectation-values shifts admit gauge-invariant decompositions into {dynamic} and {geometric} parts, related respectively to gradients of the \textit{eigenvalues} and \textit{eigenvectors} of the scattering matrix. We illustrate this general theory through a series of examples, including frequency shifts in polarized-light transmission through a time-varying waveplate (linked to the {Pancharatnam-Berry phase}), momentum shifts at spatially varying metasurfaces, optical forces, beam shifts upon reflection at a dielectric interface, and Wigner time delays in 1D scattering. This unifying framework illuminates the interplay between geometry and dynamics in wave scattering and can be applied to a broad range of physical systems.
\end{abstract}

\maketitle

%%%%%%%%%%%%%%%%%%%%%%%%%%%%%%%%%%%%%%%%%%%%%%%%%%
\section{Introduction}
%%%%%%%%%%%%%%%%%%%%%%%%%%%%%%%%%%%%%%%%%%%%%%%%%%

In 1984, Michael Berry introduced the concept of {\it geometric phase}: a fundamental addition to the usual {\it dynamic phase} in quantum (wave) evolution, which exhibits remarkable geometric properties \cite{Berry1984}. It was soon recognized that this universal concept has an extremely broad range of applications: from mechanical systems and polarization optics to Hall effects,  topological insulators, and quantum field theory \cite{Wilczek1989_book, Vinitskii1990PU, Zwanziger1990ARPC, Bhandari1997PR, Xiao2010RMP, Bliokh2015NP, Baggio2017JHEP, Cohen2019NRP, Cisowski2022RMP}. 

Initially, Berry considered the adiabatic evolution of an $n$-level quantum system characterized by a time-varying Hermitian Hamiltonian \cite{Berry1984}. In this case, the dynamic phase accumulated by an adiabatically-evolving instantaneous eigenstate is determined by an integral of the corresponding \textit{eigenvalue} of the Hamiltonian. In turn, the geometric phase is described by an integral of the {\it Berry connection}, which is proportional to the derivative of the corresponding unit \textit{eigenvector}. (The Berry connection describes local rotations and parallel transport of the eigenvectors forming a fiber bundle over the corresponding base space.) Due to adiabaticity and the time derivative involved, the geometric phase has the next order of smallness compared to the dynamic phase.
Importantly, geometric phases and Berry connections appear not only in time-varying systems but also in those with space- and momentum-dependent Hamiltonians \cite{Wilczek1989_book, Vinitskii1990PU, Zwanziger1990ARPC, Bhandari1997PR, Xiao2010RMP, Bliokh2015NP, Baggio2017JHEP, Cohen2019NRP, Cisowski2022RMP}. 

\begin{figure}[t!]
\centering
\includegraphics[width=\linewidth]{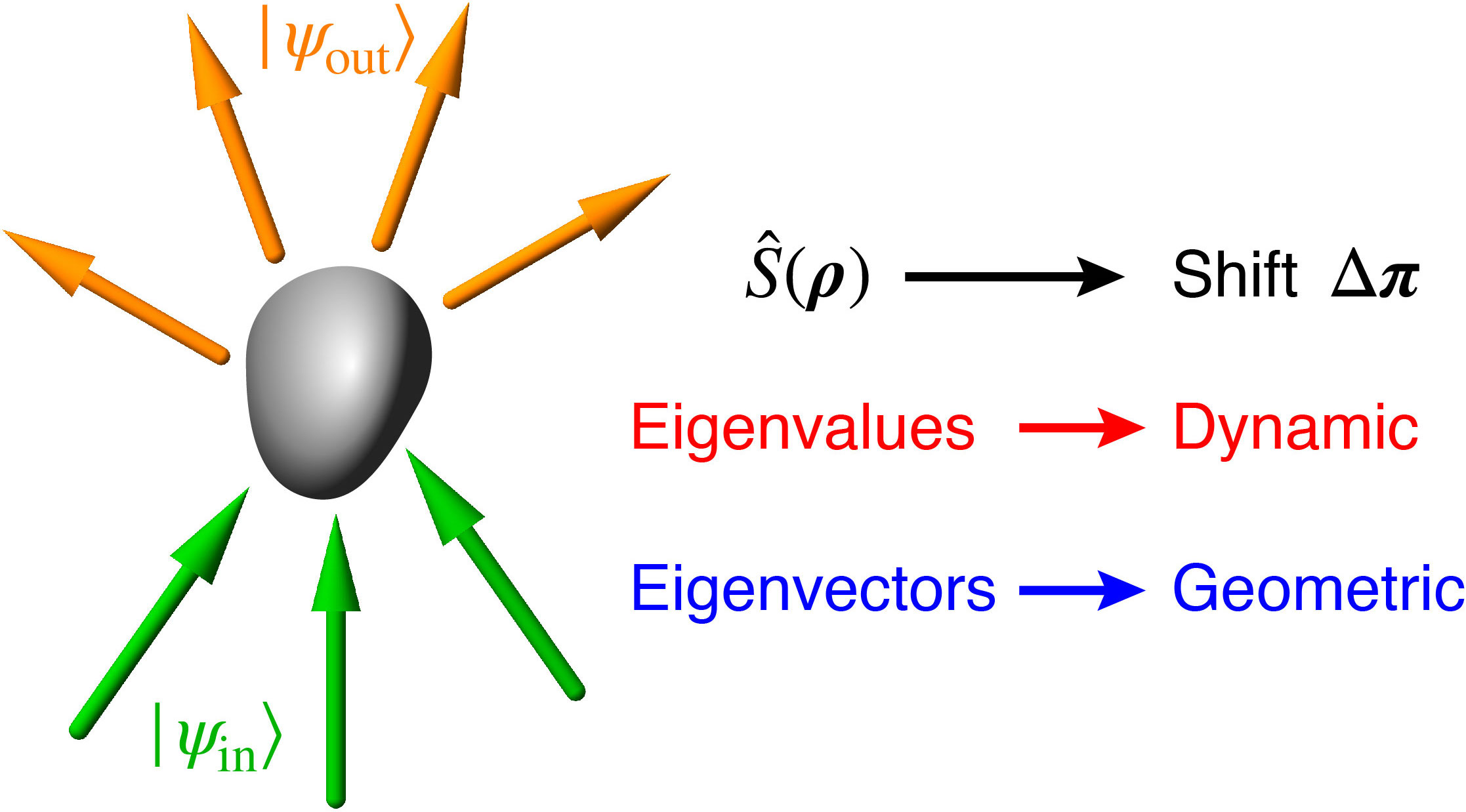}
\caption{Schematic of wave scattering described by a unitary scattering matrix $\hat{S}$, which depends on a variable ${\bm\rho}$. The scattered wave experiences a shift $\Delta {\bm \pi}$ in the expectation value of the variable conjugate to ${\bm \rho}$. This shift is described by the generalized Wigner-Smith operator (GWSO). The ${\bm\rho}$-gradients of the {\it eigenvalues} and {\it eigenvectors} of $\hat{S}$ are responsible for the {\it dynamic} and {\it geometric} parts of $\Delta {\bm \pi}$ or the GWSO, respectively.} 
\label{Fig_scattering} 
\end{figure}

In this work, we present a related decomposition of dynamic and geometric contributions, but in a principally different setting. Instead of the continuous quantum evolution of a wave state over the base space, we consider a generic {\it wave scattering} problem with varying parameters. In this context, an input wave state is directly transformed into an output state, and the unitary {\it scattering matrix} describing this transformation depends on certain basic variables (time, coordinates, momentum, etc.), as illustrated in Fig.~\ref{Fig_scattering}.
Furthermore, instead of examining the phase of the output wavefunction, we analyze the shifts in the expectation values of the variables conjugate to these basic variables. In particular, the time/position/frequency/momentum-dependent scattering produces shifts in the expectation values of frequency/momentum/time/position, respectively. 

Well-known examples of such shifts in wave-scattering problems include the Goos-H\"{a}nchen shift of an optical beam reflected from an interface characterized by a momentum-dependent reflection matrix \cite{Goos1947AP, Artmann1948AP, Bliokh2013JO}, as well as the Wigner time delay experienced by a wavepacket scattered by a frequency-dependent system \cite{Wigner1955PR, Smith1960PR, Winful2006PR, Chiao1997PO}. 
Following earlier work in mesoscopic physics \cite{Brouwer1997PRL, Brouwer1999WRM},
these phenomena have recently been generalized and extended to a broader class of wave systems and conjugate variables using the {\it generalized Wigner-Smith operator (GWSO)} framework \cite{Ambichl2017PRL, Horodynski2020NP} (see also \cite{Cao2022NP, Rachbauer2024JOSAB} for recent reviews on this subject). {In particular, this approach has been applied to optimize: wave focusing  \cite{Horodynski2020NP,PhysRevLett.134.183802}, optical forces and torques \cite{Horodynski2020NP,PhysRevA.108.023504,Orazbayev2024NP,PhysRevA.110.053515}, trapping \cite{Butaite2024SA} and cooling \cite{Hupfl2023PRL,PhysRevA.107.023112} of particles, transmission through multi-mode fibers \cite{Carpenter2015EWS,xiong_spatiotemporal_2016,Matthes2021PRX}, estimating system parameters \cite{Bouchet2021MaxInfo,bouchet_optimal_2021}, optical inverse-design \cite{Horodynski2022}, and analysis of scattering resonances \cite{PhysRevResearch.7.013299}. In all these cases, the eigenvectors of the corresponding GWSO with extreme eigenvalues determine the optimal input wave states.} 

Here we show that both the GWSOs and the corresponding expectation-value shifts in scattered waves can be unambiguously decomposed into dynamic and geometric parts. Similarly to Berry's phase decomposition, these dynamic and geometric contributions are associated with the \textit{eigenvalues} and \textit{eigenvectors} of the scattering matrix, respectively (see Fig.~\ref{Fig_scattering}). Furthermore, the geometric part is expressed via the corresponding Berry connection. However, unlike the adiabatic-evolution scenario, now the dynamic and geometric contributions have the same order of magnitude. This is because the dynamic part involves \textit{derivatives} of the eigenvalues. (This is explained by the fact that we are dealing with the eigenvalues of a unitary scattering matrix, which is analogous to the evolution operator involving an \textit{integral} of the Hamiltonian in the evolution problem.) Moreover, while the geometric phase in the adiabatic-evolution problem is gauge-invariant only for {\it cyclic} evolution in the base space, we show that the geometric and dynamic shifts in the scattering problem are gauge-invariant {\it at each point} of the base space.   

We explore applications of this general framework to various observable phenomena, mostly in optical systems. In particular, we find geometric and dynamic frequency shifts in light transmission through a time-varying waveplate, which offer a more fundamental description of the scattered light state than the conventional Pancharatnam-Berry (PB) phase \cite{Pancharatnam1956, Berry1987JMO} \blue{(for ambiguities in the PB phase description, see \cite{Vinitskii1990PU, Malykin2004PU, Simon1988PRL, Bretenaker1990PRL})}. 
In another example, the mutually-orthogonal Goos-H\"{a}nchen and Imbert-Fedorov (spin-Hall effect) shifts of an optical beam reflected from a dielectric interface \cite{Bliokh2013JO, Bliokh2015NP} are identified as the dynamic and geometric parts, respectively, of the total beam shift. 
\blue{We also describe examples of momentum shifts in light transmission through space-varying anisotropic plates (metasurfaces) and Wigner time delays in 1D wave-scattering systems.}

Thus, our work introduces a new fundamental concept, offers fresh insights and a unified perspective on a range of observable phenomena, and can have useful implications in a variety of wave systems, both classical and quantum.

%FFFFFFFFFFFFFFFFFFFFFFFFFFFFFFFFFFFFFFFFFFFFFF
%FFFFFFFFFFFFFFFFFFFFFFFFFFFFFFFFFFFFFFFFFFFFFF

%%%%%%%%%%%%%%%%%%%%%%%%%%%%%%%%%%%%%%%%%%%%%%%%%%
\section{General formalism}
%%%%%%%%%%%%%%%%%%%%%%%%%%%%%%%%%%%%%%%%%%%%%%%%%%

We consider a wave system described by an $n$-component wavefunction $| \psi \rangle$ in a flux-normalized basis, $\langle \psi | \psi \rangle =1$.
The scattering experienced by this wave is described by a unitary matrix operator $\hat{S}$ depending on some parameters (`coordinates') ${\bm \rho}$: $| \psi_{\rm out} \rangle = \hat{S}({\bm \rho})\, | \psi_{\rm in} \rangle$, see Fig.~\ref{Fig_scattering}. Assuming that the canonically-conjugate `momentum' operator is ${\bm \pi} = -i\partial/\partial {\bm \rho}$, the shift in the expectation value of ${\bm \pi}$ due to the scattering is given by \cite{Ambichl2017PRL}
\begin{equation}
\label{momentum}
\Delta {\bm \pi} = \langle {\psi}_{\rm out} | {\bm \pi} |{\psi}_{\rm out} \rangle -  \langle {\psi}_{\rm in} |{\bm \pi}| {\psi}_{\rm in}\rangle  = 
\langle {\psi}_{\rm in} |\hat{\bf Q}| {\psi}_{\rm in}\rangle\,,
\end{equation}
where $\hat{\bf Q}=-i\hat{S}^\dag  \dfrac{\partial \hat{S}}{\partial {\bm \rho}}$ is the GWSO. 
\red{For a unitary $\hat{S}$-matrix, $\hat{\bf Q}$ is Hermitian and its eigenvectors constitute an orthogonal basis with each eigenvector being associated with a wavefront that imparts a well-defined, real-valued ``momentum transfer'' between the wave and the scattering target \cite{Ambichl2017PRL, Horodynski2020NP}. %The highest and lowest possible momentum transfer is natually associated with the highest and lowest eigenvalue, respectively.
} Similarly, if the scattering matrix depends on the `momentum' ${\bm \pi}$, the shift in the expectation value of the conjugate `coordinates' ${\bm \rho} = i\partial /\partial {\bm \pi}$ is given by Eq.~\eqref{momentum}, with the substitution ${\bm \rho} \leftrightarrow {\bm \pi}$, $\partial /\partial {\bm \rho} \to - \partial /\partial {\bm \pi}$. 

The Wigner-Smith operator was originally introduced to quantify the Wigner time delay, involving a derivative with respect to frequency \cite{Wigner1955PR, Smith1960PR, Winful2006PR, Chiao1997PO}, but later it was generalized to other conjugate variables \cite{Brouwer1997PRL, Brouwer1999WRM, Ambichl2017PRL, Horodynski2020NP}. It is useful to note that the maximum shift in, e.g., the $x$-component of the `momentum' is achieved when the input wave is an eigenvector of the corresponding GWSO with the largest eigenvalue: $\hat{Q}_x |\psi_{\rm in}\rangle = q_{x,{\rm max}} |\psi_{\rm in}\rangle$ and $\Delta \pi_{x,{\rm max}} = q_{x,{\rm max}}$. 

In geometric-phase problems, the parameter-dependent diagonalization of the Hamiltonian provides an efficient method for separating the dynamic and geometric phases. Here we adopt a similar approach but apply it to the scattering matrix $\hat{S}({\bm \rho})$, which we assume to be diagonalizable via a unitary transformation
\begin{equation}
\label{U}
\hat{U}^\dag({\bm \rho}) \hat{S}({\bm \rho}) \hat{U}({\bm \rho})= \hat{S}'({\bm \rho})
= {\rm diag}\!\left(s_1({\bm \rho}),...,s_n({\bm \rho})\right)\,.
\end{equation}
Here $s_i({\bm \rho})=e^{i\alpha_i({\bm \rho})}$, $i=1,...,n$, are the unimodular eigenvalues of $\hat{S}$ \red{(which are supposed to be different from each other)}, with real-valued phases $\alpha_i$, whereas the columns of the unitary matrix $\hat{U}$ are the normalized eigenvectors of $\hat{S}$.
Making the corresponding substitution in the wavefunction, $|\psi\rangle = \hat{U} |\psi'\rangle$ and performing straightforward algebra, we re-write Eq.~\eqref{momentum} as
\begin{equation}
\label{momentum_d_g}
\Delta {\bm \pi} = 
\underbracket{\langle {\psi}'_{\rm in} |\hat{\bf Q}' | {\psi}_{\rm in}'\rangle}_{\text{dynamic}} +
\underbracket{\langle {\psi}'_{\rm out} | \hat{\bf A} | {\psi}_{\rm out}'\rangle - 
\langle {\psi}'_{\rm in} | \hat{\bf A}
| {\psi}_{\rm in}'\rangle}_{\text{geometric}}.
\end{equation}
Here $\hat{\bf Q}'=-i\hat{S}'^\dag  \dfrac{\partial \hat{S}'}{\partial {\bm \rho}}  = {\rm diag}\!\left( \dfrac{\partial \alpha_1}{\partial {\bm \rho}}, ..., \dfrac{\partial \alpha_n}{\partial {\bm \rho}}\right)$ is the diagonalized GWSO, while $\hat{\bf A}= -i\hat{U}^\dag  \dfrac{\partial \hat{U}}{\partial {\bm \rho}}$ is the Hermitian {\it Berry connection} over the ${\bm \rho}$-space. 

Equation~\eqref{momentum_d_g} consists of two contributions: (i) the expectation value of the diagonal GWSO $\hat{\bf Q}'$, which involves derivatives of the {\it eigenvalues} of the scattering matrix; and (ii) the shift in the expectation value of the Berry connection, which is associated with derivatives of the {\it eigenvectors} of the scattering matrix. These two contributions determine, respectively, the {\it dynamic} ($\Delta {\bm \pi}_d$) and {\it geometric} ($\Delta {\bm \pi}_g$) parts of the total momentum shift \eqref{momentum}, see Fig.~\ref{Fig_scattering}. 

Transforming Eq.~\eqref{momentum_d_g} back to the original basis, $|\psi'\rangle = \hat{U}^\dag |\psi\rangle$, and comparing it with Eq.~\eqref{momentum}, we find that the GWSO $\hat{\bf Q}$ can also be decomposed into dynamic and geometric (both Hermitian) parts, $\hat{\bf Q} = \hat{\bf Q}_d + \hat{\bf Q}_g$: 
\begin{equation}
\label{Qd-Qg}
\hat{\bf Q}_d = 
\hat{U}\hat{\bf Q}'\hat{U}^\dag,\quad
\hat{\bf Q}_g =  
\hat{U}\! \left( \hat{S}^{\prime\dag} \hat{\bf A} \hat{S}' -  \hat{\bf A} \right)\!\hat{U}^\dag. 
\end{equation}

Equations~\eqref{momentum_d_g} and \eqref{Qd-Qg} are the key general results of this work. 
They establish a geometric-dynamic decomposition of the shift in the conjugate variable (`momentum'), as well as of the associated GWSO.
Note that the geometric part of the GWSO is traceless: ${\rm tr}(\hat{\bf Q}_g)=0$ \red{(this follows from the additivity of the trace and its invariance with respect to unitary transformations)}. Considering, e.g., the $x$-components of the shifts, it is easy to show that if the input wave is an eigenvector of the dynamic GWSO, $\hat{Q}_{d,x} |\psi_{\rm in}\rangle = q_{d,x,i} |\psi_{\rm in}\rangle$, then the geometric shift vanishes: $\Delta { \pi}_{g,x} = \langle \psi_{\rm in} |\hat{Q}_{g,x} |\psi_{\rm in}\rangle =0$. 
In particular, the maximum dynamic shift $\Delta {\pi}_{d,x} = q_{d,x, {\rm max}}$ is necessarily accompanied by the vanishing geometric shift. 
In turn, the maximum total shift cannot exceed the sum of the maximum dynamic and geometric shifts: $\Delta {\pi}_{x,{\rm max}} \leq \Delta {\pi}_{d,x,{\rm max}} + \Delta {\pi}_{g,x,{\rm max}}$. This follows from Weyl's inequality for eigenvalues of a sum of Hermitian matrices.   

Importantly, the geometric-dynamic decomposition \eqref{momentum_d_g} and \eqref{Qd-Qg} is {\it gauge-invariant}. The diagonalizing transformation \eqref{U} \red{has a gauge freedom: it} is defined up to a diagonal gauge transformation: $\hat{U}({\bm \rho}) \to \hat{U}({\bm \rho})\hat{G}({\bm \rho})$, where $\hat{G}({\bm \rho})={\rm diag}\!\left( e^{i\gamma_1({\bm \rho})},...,e^{i\gamma_n({\bm \rho})} \right)$. This transformation leaves the diagonalized scattering matrix and the GWSO invariant, $\hat{S}' \to \hat{S}'$, $\hat{\bf Q}' \to \hat{\bf Q}'$, but modifies the Berry connection: $\hat{\bf A} \to \hat{G}^\dag \hat{\bf A} \hat{G} - i \hat{G}^\dag\dfrac{\partial \hat{G}}{\partial {\bm \rho}}$. 
Nonetheless, straightforward calculation shows that the dynamic and geometric `momentum' shifts, $\Delta {\bm \pi}_d$ and $\Delta {\bm \pi}_g$, as well as the corresponding GWSOs, remain gauge-invariant: $\hat{\bf Q}_d \to \hat{\bf Q}_d$, $\hat{\bf Q}_g \to \hat{\bf Q}_g$.
\red{(This follows from the relations $\hat{\bf Q}_d \to \hat{U}\hat{G}\hat{\bf Q}' \hat{G}^\dag \hat{U}^\dag = \hat{U}\hat{\bf Q}' \hat{U}^\dag = \hat{\bf Q}_d$ and $\hat{\bf Q}_g = \hat{\bf Q} - \hat{\bf Q}_d$.)
The gauge invariance} indicates that the geometric-dynamic decomposition \eqref{momentum_d_g} and \eqref{Qd-Qg} is physically meaningful and observable.

%%%%%%%%%%%%%%%%%%%%%%%%%%%%%%%%%%%%%%%%%%%%%%%%%%
\section{Examples}
%%%%%%%%%%%%%%%%%%%%%%%%%%%%%%%%%%%%%%%%%%%%%%%%%%

We now apply the general framework to several examples of wave systems that exhibit dynamic and geometric shifts in different conjugate variables.

%%%%%%%%%%%%%%%%%%%%%%%%%%%%%%%%%%%%%%%%%%%%%%%%%%
\subsection{Frequency shifts at time-varying waveplate}
%%%%%%%%%%%%%%%%%%%%%%%%%%%%%%%%%%%%%%%%%%%%%%%%%%

\begin{figure}[t!]
\centering
\includegraphics[width=0.9\linewidth]{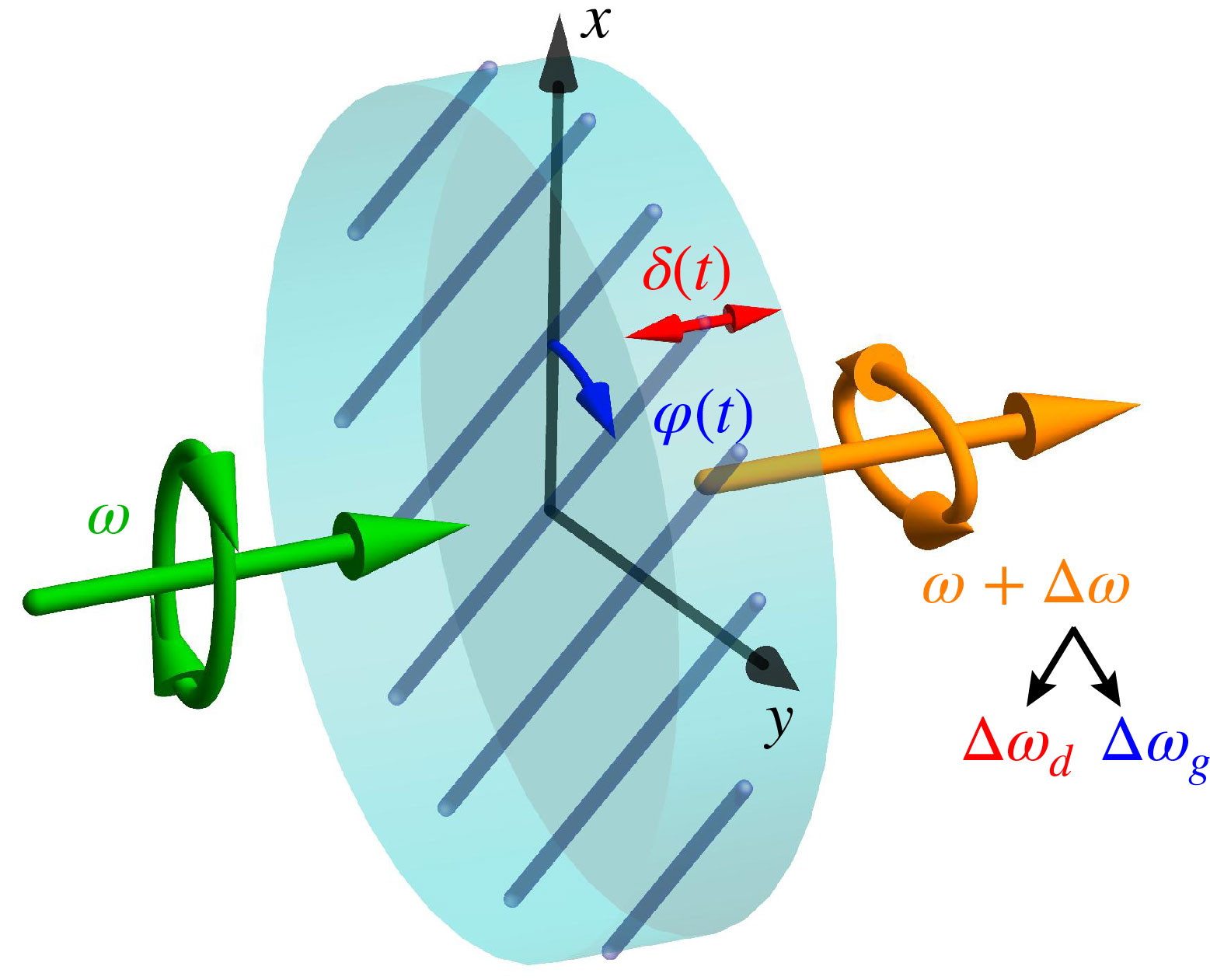}
\caption{Transmission of polarized light (without reflection) through an anisotropic waveplate with time-varying retardation between orthogonal polarizations, $\delta(t)$, and orientation of the \red{optical} axis, $\varphi(t)$, described by the scattering matrix \eqref{S_plate}. The input and output polarization states are shown by elliptical arrows. The transmitted light undergoes a frequency shift $\Delta \omega$. It includes dynamic and geometric contributions \eqref{Delta_omega} described by the corresponding parts of the GWSO, Eqs.~\eqref{Qd-Qg_plate}, and originating from the time derivatives of the parameters $\delta$ and $\varphi$, respectively.} 
\label{Fig_plate} 
\end{figure}

We start with examining the transmission of polarized light through a uniaxial anisotropic plate (waveplate), as shown in Fig.~\ref{Fig_plate}. This problem represents reflection-less two-channel scattering, corresponding to the transmission of two polarization modes. These modes can be described by the wavefunction (normalized polarization Jones vector) in the circular-polarizations basis  $|\psi\rangle = \left( \begin{matrix} e^+ \\ e^- \end{matrix} \right)$, $|e^+|^2 + |e^-|^2=1$, with the corresponding scattering (Jones) matrix \cite{Bomzon2002OL, Hasman2005PO, Bliokh2015NP}:
\begin{equation}
\hat{S} = \left( \begin{matrix} \cos\! \frac{\delta}{2} & i e^{-2i\varphi} \sin\! \frac{\delta}{2}  \\ i e^{2i\varphi} \sin\! \frac{\delta}{2} & \cos\! \frac{\delta}{2}  \end{matrix} \right).
\label{S_plate}
\end{equation}
Here $\varphi$ is the angle of the \red{optical (anisotropy)} axis of the plate in the $(x,y)$-plane (assuming light propagates along the $z$-axis), and $\delta$ is the retardation phase between the orthogonal linear polarizations aligned with the \red{principal optical} axes, which is determined by the \red{optical} thickness of the anisotropic plate.

%FFFFFFFFFFFFFFFFFFFFFFFFFFFFFFFFFFFFFFFFFFFFFF
%FFFFFFFFFFFFFFFFFFFFFFFFFFFFFFFFFFFFFFFFFFFFFF

Let the waveplate parameters vary in time: $\delta=\delta(t)$ and $\varphi = \varphi (t)$.
This generates a shift in frequency (associated with the operator $i\partial/\partial t$), which is described by the GWSO $\hat{Q} = i\hat{S}^\dag \dfrac{\partial \hat{S}}{\partial t}$. The matrix \eqref{S_plate} can be diagonalized by the unitary transformation
\begin{equation}
\hat{U} = \frac{1}{\sqrt{2}}\!\left( \begin{matrix} e^{-i\varphi} &  -i e^{-i\varphi} \\ 
e^{i\varphi} & i e^{i\varphi} \end{matrix} \right)\!,~
\hat{S}' = {\rm diag}\!\left( e^{i\delta/2}, e^{-i\delta/2} \right)\!.
\label{U_plate}
\end{equation}
The corresponding Berry connection is then given by 
\begin{equation}
\hat{A} =  i \hat{U}^\dag \dfrac{\partial \hat{U}}{\partial t} = \!\left( \begin{matrix} 0 & -i \\ i & 0\end{matrix} \right)\! \dfrac{\partial \varphi}{\partial t}\,.
\label{A_plate}
\end{equation}

Substituting Eqs.~\eqref{U_plate} and \eqref{A_plate} into general Eqs.~\eqref{Qd-Qg}, we obtain the dynamical and geometric part of the GWSO:
\begin{align}
\hat{Q}_d & =-\frac{1}{2}\!\left( \begin{matrix} 0 & e^{-2i\varphi} \\ e^{2i\varphi} & 0\end{matrix} \right)\!\frac{\partial \delta}{\partial t}\,,\nonumber \\
\hat{Q}_g & =\left( \begin{matrix} -1 +\cos\delta & i e^{-2i\varphi} \sin\delta \\ -i  e^{2i\varphi} \sin\delta & 1 -\cos\delta\end{matrix} \right)\!\frac{\partial \varphi}{\partial t}\,. 
\label{Qd-Qg_plate}
\end{align}
One can see that the eigenvalues and eigenvectors (forming the diagonalization matrix $\hat{U}$) of the scattering matrix $\hat{S}$ are determined by the parameters $\delta$ and $\varphi$, respectively. Correspondingly, the dynamic and geometric parts of the GWSO are proportional to the time derivatives of these parameters  
\footnote{It is worth noticing that our association of the $\delta(t)$ and $\varphi(t)$ variations with dynamic and geometric effects, respectively, is consistent with the original Berry-phase formalism for the spin 1/2 evolution in a time-varying magnetic field \cite{Berry1984}. Indeed, expressing the scattering matrix \eqref{S_plate} as $\hat{S}=\exp(i\hat{H})$, the effective Hamiltonian has the form $\hat{H} = {\bf B}\cdot \hat{\bm \sigma}$, where $\hat{\bm \sigma}$ is the vector of Pauli matrices, whereas the effective magnetic field reads ${\bf B}=\delta\,(\cos2\varphi,\sin2\varphi,0)$. Thus, the magnitude of the field, responsible for the dynamic phase, is provided by $\delta$, while the direction of the field, responsible for the geometric phase, is governed by $\varphi$.}.

According to Eq.~\eqref{momentum_d_g}, the expectation values of the GWSOs \eqref{Qd-Qg_plate} with the input wavefunction $|\psi_{\rm in}\rangle$ yield the dynamic and geometric frequency shifts in the transmitted light, Fig.~\ref{Fig_plate}:
\begin{align}
\label{Delta_omega}
\Delta \omega_d & =-\frac{1}{2}\!\left( \tau \cos 2\varphi + \chi \sin 2\varphi \right)\!\frac{\partial \delta}{\partial t}\,, \\
\Delta \omega_g & = -\left[ \sigma (1-\cos\delta) + \chi \cos 2\varphi \sin\delta - \tau \sin 2\varphi \sin\delta\right]\!\frac{\partial \varphi}{\partial t}\,. \nonumber
\end{align}
Here $\tau = 2 {\rm Re}(e^{+*}_{\rm in} e^-_{\rm in})$, $\chi = 2 {\rm Im}(e^{+*}_{\rm in}e^-_{\rm in})$, $\sigma = |e^{+}_{\rm in}|^2 - |e^-_{\rm in}|^2$, $\tau^2+\chi^2+\sigma^2 =1$, are the Stokes parameters of the incoming light that map its polarization state onto the unit Poincar\'{e} sphere \cite{Azzam_book}. 

The time integral of the frequency shift yields the total phase shift accumulated in the transmitted light (with respect to the incident light) over a given time interval: $\Delta \Phi = - \int \Delta \omega\, dt$. 
It also determines the total energy exchange, i.e., the {\it work} produced by the waveplate on the light \cite{Allen1966AJP, Bretenaker1990PRL, Rakich2009OE, Kuang2025}: $W = - (cI/\omega) \Delta\Phi$, \blue{where the factor $cI/\omega$ (with $I$ denoting the electromagnetic energy density per unit $z$-length in the incident beam of frequency $\omega$) equals $\hbar$ times the number of photons transmitted through the waveplate per unit time.} 

Equations~\eqref{Delta_omega} have several important implications. Although the dynamic and geometric parts contribute to the total observable frequency shift $\Delta\omega$ and the integral phase shift $\Delta\Phi$, they arise from qualitatively distinct physical origins.

First, the parameter $\varphi$ is a \textit{cyclic} variable corresponding to the \textit{orientation} of the waveplate. Increasing it by $\pi$ (with a fixed $\delta$) returns the system to its initial state, but with a nonzero accumulated geometric phase $\Delta\Phi_g = - \int \Delta\omega_g\, dt = \pi\sigma(1-\cos\delta)$. In contrast, increasing the parameter $\delta$ by $4\pi$ (with a fixed $\varphi$) restores the original form of the Jones matrix \eqref{S_plate}, but this evolution corresponds to a physical change in the \red{optical} \textit{thickness} of the plate -- a \textit{non-cyclic} variable. 
A truly cyclic evolution in $\delta$, returning the plate to its initial state, results in zero dynamic phase accumulation: $\Delta\Phi_d =0$. 

Second, the geometric frequency shift \eqref{Delta_omega} is directly related to the coupling between the \textit{angular momentum} of light and the rotations of the system. This mechanism underlies geometric phases in a variety of systems \cite{Lipson1990OL, Opat1991AJP, Khein1993AJP, Serebrennikov2006PRB, Bliokh2015NP, Bliokh2008PRL, Bliokh2009JOA}. Specifically, the $z$-component of the normalized spin angular momentum of the incoming light is determined by the third Stokes parameter $\sigma$, which quantifies the degree of circular polarization. Calculating the Stokes parameters of the transmitted light, $(\tau',\chi',\sigma')$ [see Appendix~A], we find that the geometric frequency shift \eqref{Delta_omega} can be written as 
\begin{equation}
\Delta\omega_g = (\sigma' -\sigma)\frac{\partial\varphi}{\partial t}\,.
\label{Coriolis}
\end{equation}
This is the standard form of the \textit{Coriolis}  \cite{Mashhoon1988PRL, Opat1991AJP} or \textit{angular Doppler} \cite{Garetz1979OC, Garetz1981JOSA, Bialynicki-Birula1997PRL, Emile2023AP} effect, which underlies the appearance of geometric phases \cite{Lipson1990OL, Serebrennikov2006PRB, Bliokh2015NP, Bliokh2008PRL, Bliokh2009JOA}. \blue{In contrast, considering the dynamic frequency shifts \eqref{Delta_omega}, evaluated for $\tau=\pm 1$ and $\varphi=0$, or for $\chi=\pm 1$ and $\varphi =\pi/4$, one can see that it is equal to the minus time derivative of the anisotropy-induced \textit{optical path} addition $\pm\delta/2$ for polarizations aligned with the fast and slow axes of the waveplate. This agrees with the common association of the optical path with the {dynamic phase} of light  \cite{Wilczek1989_book, Vinitskii1990PU, Bhandari1997PR}.}

Notably, the deep connection between rotations and geometric effects extends to the wave scattering of an {\it arbitrary rotating body} (assuming the absence of inhomogeneities or other bodies in the surrounding space). Indeed, rotations of the scattering object do not change the eigenvalues of the scattering matrix, and can only affect its eigenvectors. Therefore, any rotations can produce only \textit{geometric} GWSO and associated conjugate-variable shifts.

%%%%%%%%%%%%%%%%%%%%%%%%%%%%%%%%%%%%%%%%%%%%%%%%%%
\subsection{Relation to the Pancharatnam-Berry phase}
%%%%%%%%%%%%%%%%%%%%%%%%%%%%%%%%%%%%%%%%%%%%%%%%%%

The geometric and dynamic frequency shifts \eqref{Delta_omega} at a time-varying waveplate exhibit rather nontrivial interconnections with the PB phase, which is widely recognized as the geometric phase associated with polarization evolution \cite{Pancharatnam1956, Berry1987JMO, Vinitskii1990PU, Bhandari1997PR}. We recall that for a cyclic polarization evolution, the PB phase is given by minus half the solid angle $\Omega$ embraced by the evolving Stokes vector $\vec{\Sigma}=(\tau,\chi,\sigma)$ on the unit Poincar\'{e} sphere: $\Phi_{PB} = - \Omega/2$. 
\red{For the generic time-varying waveplate and a generic incident polarization state, the Poincar\'{e}-sphere calculations are rather cumbersome. Therefore, we examine particular cases of circular and linear incoming polarizations, which are sufficient for our purposes.}

\begin{figure}[t!]
\centering
\includegraphics[width=0.8\linewidth]{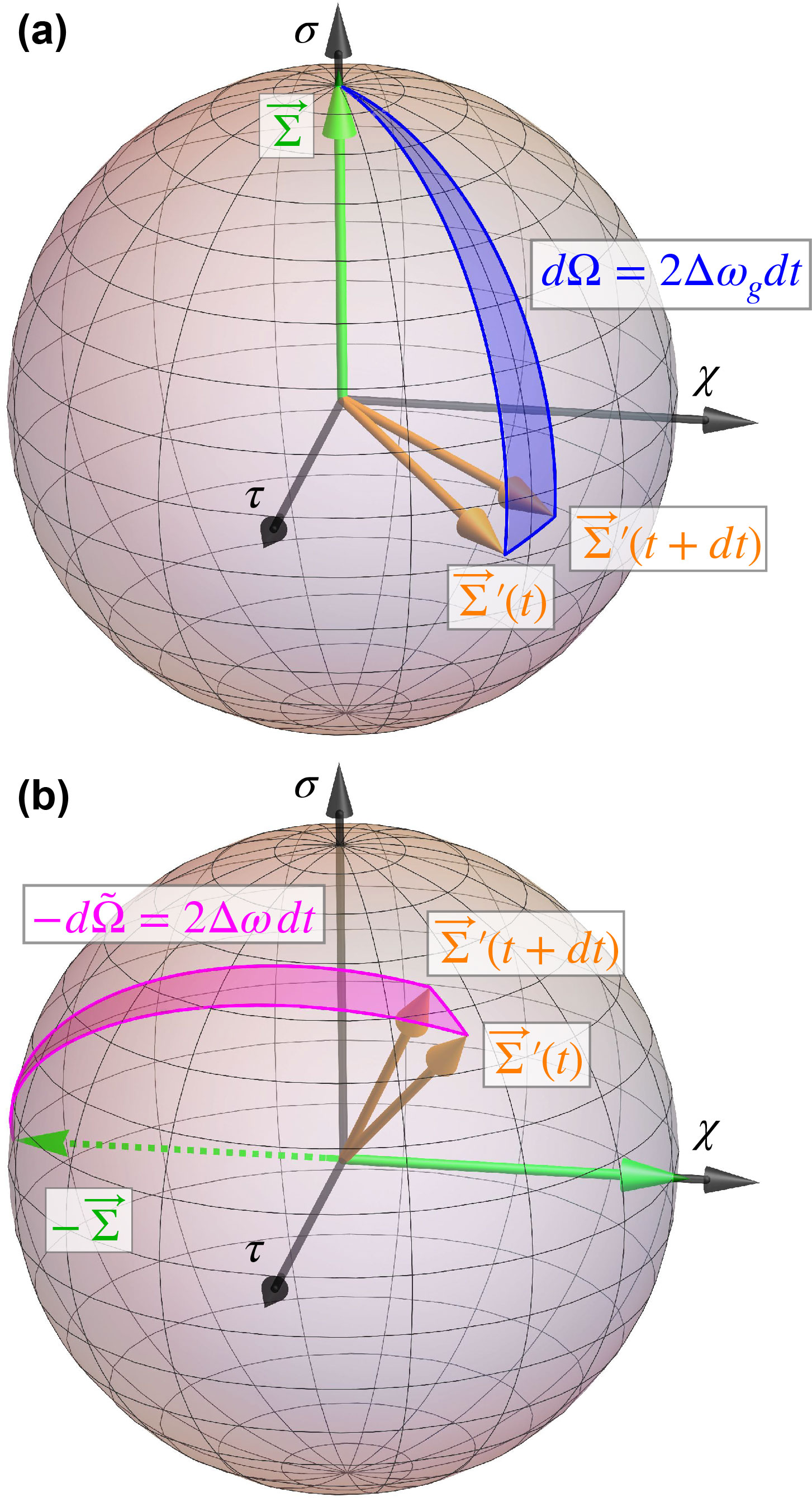}
\caption{Connection between frequency shifts \eqref{Delta_omega} and the Pancharatnam-Berry (PB) phase on the Poincar\'{e} sphere. 
(a) For circularly-polarized incident light ($\sigma=1$ here), the infinitesimal geometric phase shift $\Delta\omega_g dt$ is related to the infinitesimal solid angle $d\Omega$ formed by the Stokes vectors of the incident light, $\vec{\Sigma}$, transmitted light, $\vec{\Sigma}'(t)$, and $\vec{\Sigma}'(t+dt)$, via the corresponding PB phase, Eq.~\eqref{PB_1}. The dynamic shift vanishes in this case: $\Delta\omega_d=0$. (b) For linearly-polarized incident light ($\chi=1$ here), the infinitesimal \textit{total} phase shift $\Delta\omega\, dt$ is related (with a minus sign) to the modified solid angle $d\tilde{\Omega}$ formed by the Stokes vectors $-\vec{\Sigma}$, $\vec{\Sigma}'(t)$, and $\vec{\Sigma}'(t+dt)$, via the modified PB phase, Eq.~\eqref{PB_2}. In both cases (a) and (b), for a cyclic evolution, when the output Stokes vector $\vec{\Sigma}'(t)$ traces a closed loop on the sphere, the total phase accumulated in the transmitted light is given by the integral PB expression \eqref{PB_cyclic} involving the solid angle enclosed by the loop.} 
\label{Fig_sphere} 
\end{figure}

Consider first the case of circularly-polarized incident light: $\sigma=\pm 1$, $\tau=\chi=0$. Then, the dynamic frequency shift \eqref{Delta_omega} vanishes, $\Delta\omega_d=0$, whereas the geometric frequency shift is directly related to the PB phase [see Appendix~A]: 
\begin{equation}
\Delta\omega\, dt=\Delta\omega_g dt= - d\Phi_{PB} = \frac{1}{2} d\Omega\,.
\label{PB_1}
\end{equation}
Here the infinitesimal solid angle $d\Omega$ is formed by: (i) the Stokes vector of the incident light, $\vec{\Sigma}$, (ii) the Stokes vector of the transmitted light $\vec{\Sigma}'(t)$, (iii) $\vec{\Sigma}'(t + dt)$, and geodesic lines on the Poincar\'{e} sphere connecting these points, see Fig.~\ref{Fig_sphere}(a). 
Equation~\eqref{PB_1} also holds for linearly-polarized incident light, circularly-polarized transmitted light, and $\varphi(t)$-rotations of the waveplate with $\delta={\rm const}$. These configurations, involving incident circular or linear polarizations and rotating quarter-wave or half-wave plates, were employed in most of the key experimental demonstrations of the PB phase and its associated frequency shifts \cite{Garetz1979OC, Bhandari1988PRL, Chyba1988OL, Simon1988PRL, Bretenaker1990PRL, Mashhoon1998PLA}. 

%FFFFFFFFFFFFFFFFFFFFFFFFFFFFFFFFFFFFFFFFFFFFFF
%FFFFFFFFFFFFFFFFFFFFFFFFFFFFFFFFFFFFFFFFFFFFFF

However, in the more general case with non-circular incident polarization combined with $\delta(t)$ variations of the plate, leading to a nonzero dynamic frequency shift, the relation \eqref{PB_1} fails. In particular, for linearly-polarized incident light with $\tau=\pm 1$ or $\chi=\pm 1$, the \textit{total} frequency shift produces the phase equivalent to a \textit{modified} form of the PB phase [see Appendix~A]: 
\begin{equation}
\Delta\omega\, dt = (\Delta\omega_d + \Delta\omega_g) dt= - d\tilde{\Phi}_{PB} = - \frac{1}{2} d\tilde{\Omega}\,.
\label{PB_2}
\end{equation}
Here the solid angle $d\tilde{\Omega}$ is formed by the Stokes vectors (i) $-\vec{\Sigma}$, (ii) $\vec{\Sigma}'(t)$, and (iii) $\vec{\Sigma}'(t+dt)$, as shown in Fig.~\ref{Fig_sphere}(b). 

For a cyclic evolution of the output Stokes vector $\vec{\Sigma}'(t)$, the integrals of Eqs.~\eqref{PB_1} and \eqref{PB_2} yield the same value of the total phase modulo $2\pi$:
\begin{equation}
\Delta\Phi~{\rm mod}~2\pi=-\int\Delta\omega\, dt ~{\rm mod}~2\pi= -\frac{1}{2} {\Omega} = \Phi_{PB}\,,
\label{PB_cyclic}
\end{equation}
where $\Omega$ is the solid angle encompassed by the closed trajectory of $\vec{\Sigma}'(t)$. Equation~\eqref{PB_cyclic} is consistent with the standard definition of the PB phase and reveals that, surprisingly, the \textit{PB phase incorporates both dynamic and geometric contributions} to the frequency shift. Moreover, Eqs.~\eqref{PB_1} and \eqref{PB_2} show that for non-cyclic evolution, \textit{there is no universal differential PB form on the Poincar\'{e} sphere corresponding to the observable frequency shift}.
\blue{Remarkably, our approach is free of these peculiarities and ambiguities of the PB phase, and provides a gauge-invariant decomposition of dynamic and geometric effects, Eqs.~\eqref{Delta_omega}, at a more fundamental level, not requiring integration over a cyclic evolution.}

Two observations can explain the limitations of the PB-phase approach in the system under consideration. First, light transmission through a time-varying optical element is not a problem with continuous evolution of time-varying instantaneous eigenmodes, for which the geometric-phase approach was originally introduced \cite{Berry1984}. It is rather a time-dependent \textit{scattering} problem, where the input and output states are \red{\textit{not}} eigenmodes of the scattering matrix. 
%(Note also the inconsistency of the PB-phase approach to the polarization evolution upon propagation of light in anisotropic media \cite{Malykin2004PU}. This evolution is a result of interference of different eigenmodes.) 
\blue{(See also \cite{Malykin2004PU} for a discussion of inconsistencies in applying the PB-phase approach to polarization evolution during light propagation in stationary anisotropic media, where the evolution arises from interference between different eigenmodes.)}
Second, the PB formalism, \red{dealing with solid angles and geodesic lines,} treats the Poincar\'{e} sphere as an \textit{isotropic} phase space, which is mathematically elegant and convenient in many problems. However, from a physical standpoint, the Poincar\'{e} sphere is actually \textit{anisotropic}. Indeed, only its $\sigma$ axis corresponds to the spin angular momentum \cite{Berry1998SPIE} and therefore contributes to frequency or phase shifts via the Coriolis or angular-Doppler effect \eqref{Coriolis}. In turn, the $\tau$ and $\chi$ axes are unrelated to the angular momentum and contribute to the frequency and phase shifts via different mechanisms.

Finally, we note that the total phase shift $\Delta\Phi = -\int \Delta\omega\, dt$ can be expressed as a contour integral of a tangent vector field over the unit sphere parametrized by the spherical angles $(\delta,2\varphi)$ [see Appendix~B]. This field depends on the Stokes parameters of the incident light and takes the form of the Pancharatnam-like connection (corresponding to half the solid angle on the sphere) only for circular input polarization: $\sigma=\pm 1$, $\tau=\chi=0$. Thus, only in this special case does the observable frequency shift $\Delta\omega = \Delta\omega_g$ admit a simple geometric interpretations on both the Poincar\'{e} and $(\delta,2\varphi)$ spheres.

%%%%%%%%%%%%%%%%%%%%%%%%%%%%%%%%%%%%%%%%%%%%%%%%%%
\subsection{Space-varying metasurfaces and optical forces}
%%%%%%%%%%%%%%%%%%%%%%%%%%%%%%%%%%%%%%%%%%%%%%%%%%

\begin{figure}[t!]
\centering
\includegraphics[width=0.75\linewidth]{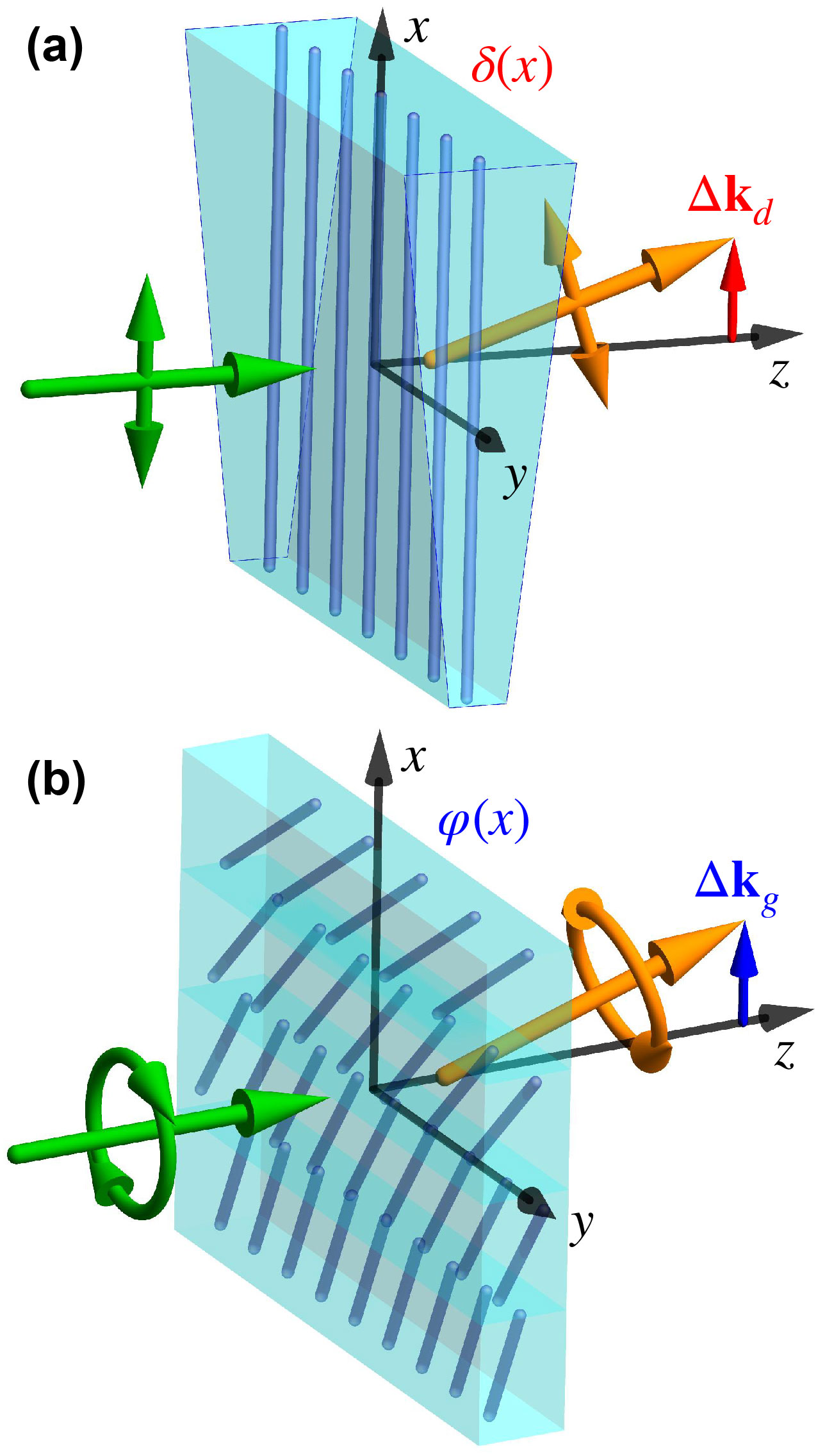}
\caption{Transmission of light through a waveplate with spatially varying parameters (a metasurface). (a) Similarly to the time-varying phase plate, Fig.~\ref{Fig_plate}, spatial variations of the phase retardation (plate thickness), $\delta({\bf r})$, produce a dynamic wavevector shift $\Delta{\bf k}_d$. Shown is the case of linear input polarization ($\tau=1$). (b) Spatial variations of the \red{optical}-axis orientation, $\varphi({\bf r})$, generate a geometric wavevector shift $\Delta{\bf k}_g$. Shown is the case of circular input polarization ($\sigma=1$) and half-wave retardation ($\delta=\pi$). These polarization-dependent wavevector shifts are described by Eqs.~\eqref{Delta_omega} and \eqref{Qd-Qg_plate} with the substitutions $\Delta\omega \to \Delta{\bf k}$ and $\partial/\partial t \to - \partial/\partial{\bf r}$.} 
\label{Fig_metasurface} 
\end{figure}

The previous example of a time-varying waveplate can be extended to a waveplate with {\it spatially} varying retardation and \red{optical}-axis orientation: $\delta=\delta({\bf r})$ and $\varphi=\varphi({\bf r})$, see Fig.~\ref{Fig_metasurface}. This corresponds to a {\it metasurface} \cite{Hasman2005PO, Yu2014NM} with varying thickness, which produces {\it wavevector} shifts in the transmitted light. Equations~\eqref{S_plate}--\eqref{Delta_omega} remain applicable under the substitutions $\partial/\partial t \to - \partial /\partial {\bf r}$ and $\Delta\omega \to \Delta {\bf k}$.

%FFFFFFFFFFFFFFFFFFFFFFFFFFFFFFFFFFFFFFFFFFFFFF
%FFFFFFFFFFFFFFFFFFFFFFFFFFFFFFFFFFFFFFFFFFFFFF

In this manner, the dynamic wavevector shift $\Delta{\bf k}_d \propto \partial \delta /\partial {\bf r}$ corresponds to light deflection due to the spatial gradient of thickness, like in a prism, see Fig.~\ref{Fig_metasurface}(a), while the geometric shift $\Delta{\bf k}_g \propto \partial \varphi /\partial {\bf r}$ is employed in metasurfaces for geometric-phase-based shaping of transmitted light \cite{Hasman2005PO, Bliokh2015NP}, see Fig.~\ref{Fig_metasurface}(b). In particular, \red{a half-wave metasurface ($\delta=\pi$) illuminated by circularly-polarized light ($\sigma=\pm 1$) provides the optimal conditions for the geometric momentum shift; for linear dependence $\varphi = q x$} it produces a spin-dependent deflection of the transmitted light: $\Delta k_x = 2 \sigma q$ \cite{Bomzon2002OL, Hasman2005PO, Yu2011S, Bliokh2015NP}. Similarly, if the \red{optical}-axis orientation varies with the azimuthal angle $\phi$ in the $(x,y)$ plane, $\varphi=q\phi$, $q=0,\pm 1/2, \pm 1, ...$, this produces a shift in the azimuthal (i.e., {\it orbital angular momentum}) quantum number of the transmitted light: $\Delta\ell = 2\sigma q$. This mechanism underlies the action of q-plates, which generate optical vortices with topological charge $\Delta\ell$ \cite{Biener2002OL, Marrucci2006PRL, Hasman2005PO, Marrucci2011JO, Bliokh2015NP}.

In addition, spatially varying scattering has interesting implications for {\it optical forces}. Consider light scattering from an {\it arbitrary body} (particle), with the center located at position ${\bf r}_0$, in an otherwise homogeneous and isotropic medium. Then, the momentum transfer to the particle, $\Delta {\bf p}_0$, and hence the optical force, is characterized by the GWSO $\hat{\bf Q}=-i\hat{S}^\dag  \dfrac{\partial \hat{S}}{\partial {\bf r}_0}$ \cite{Horodynski2020NP, Hupfl2023PRL, Orazbayev2024NP}. Since translations of the particle do not affect the eigenvalues of the scattering matrix, the resulting optical force has a purely {\it geometric} origin. Moreover, similarly to the work $W\propto \int \Delta \omega\, dt$ in the time-varying problem, the optical work on the particle in the space-varying case is given by $W\propto \int\Delta{\bf p}_0 \cdot d{\bf r}_0$ \cite{Rakich2009OE, Kuang2025}. 
Since the GWSO $\hat{\bf Q}$ does not have the form of the gradient of a scalar potential, transporting the particle along a closed contour generally results in nonzero net work: $\oint \Delta{\bf p}_0 \cdot d{\bf r}_0 \neq 0$. This indicates that the optical force is generally {\it non-conservative} \cite{Berry2013JPA, Sukhov2017RPP, Toftul2024}. A simple example is provided by an isotropic particle moving along a circle around the center of an optical vortex. In this case, the radiation-pressure force has a constant azimuthal component, resulting in nonzero work over the closed trajectory \cite{Toftul2024, ONeil2002PRL, Garces-Chavez2003PRL, Curtis2003PRL}. 

%%%%%%%%%%%%%%%%%%%%%%%%%%%%%%%%%%%%%%%%%%%%%%%%%%
\subsection{Beam shifts at an interface}
%%%%%%%%%%%%%%%%%%%%%%%%%%%%%%%%%%%%%%%%%%%%%%%%%%

We now consider a scenario involving a momentum-dependent scattering matrix. A simple example is provided by the reflection of a paraxial Gaussian-like optical beam from a planar dielectric interface. The Fresnel reflection coefficients for plane waves in the incident beam depend on the direction of the wavevector ${\bf k}$, 
and, hence, the scattering martix is ${\bf k}$-dependent \cite{Bliokh2013JO}. For simplicity, we focus on the case of total internal reflection, with no transmission and two reflection channels corresponding to the two polarization states. 

\begin{figure}[t!]
\centering
\includegraphics[width=0.85\linewidth]{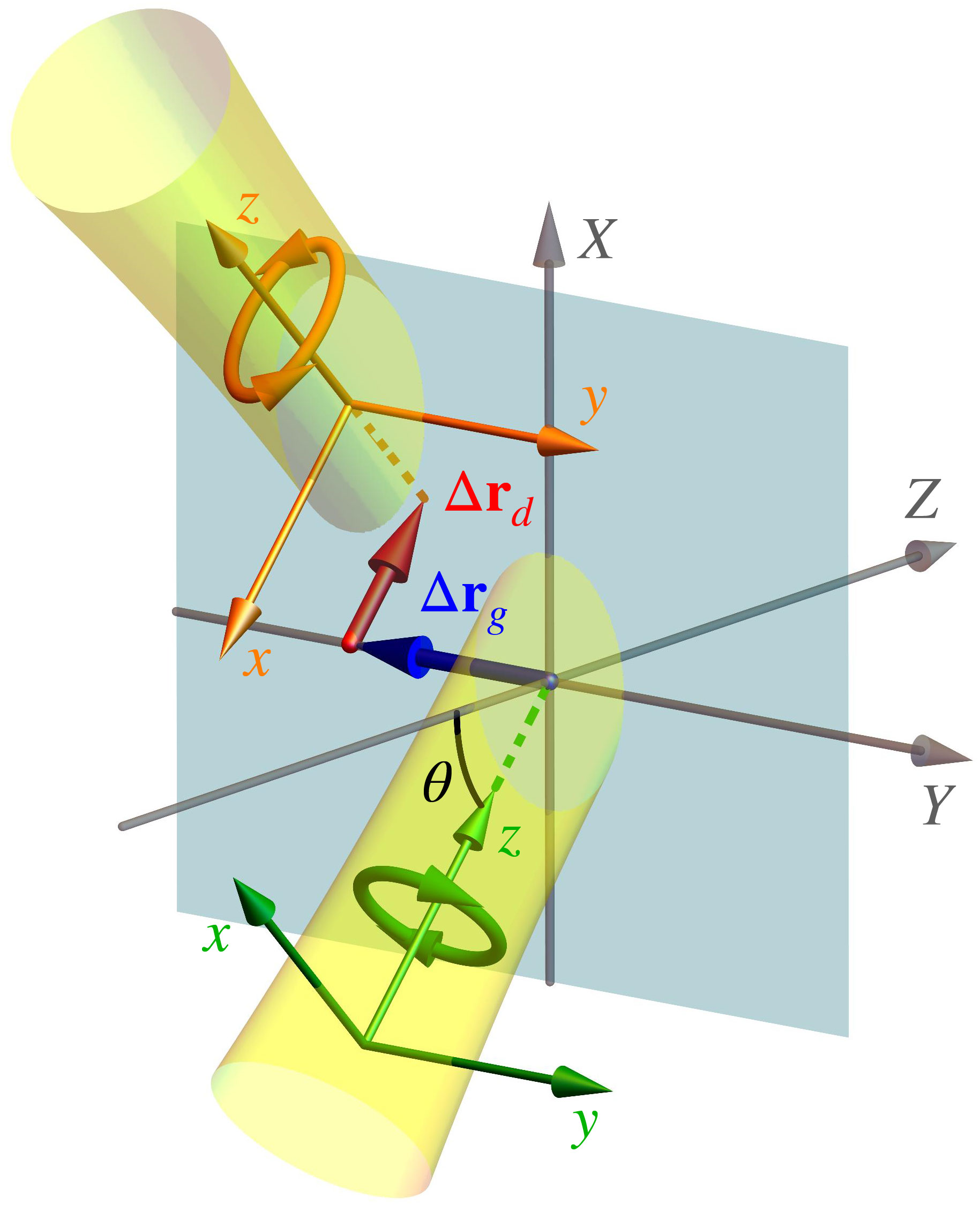}
\caption{Total internal reflection of a Gaussian-like beam at a dielectric interface, described by the scattering matrix $\hat{S}({\bf k}_\perp)$, Eq.~\eqref{S_beam} (see explanations in the text). The $x$- and $y$-directed shifts of the reflected beam, $\Delta{\bf r}_d$ and $\Delta{\bf r}_g$, correspond to the Goos-H\"{a}nchen and Imbert-Fedorov (spin-Hall effect) shifts, respectively, and are described by the dynamical and geometric parts of the GWSO, Eqs.~\eqref{Qd-Qg_beam} and \eqref{Delta_R}.} 
\label{Fig_beam} 
\end{figure}

The geometry of the problem is shown in Fig.~\ref{Fig_beam}: the interface lies in the $(X,Y)$ plane and the beam propagates at the angle $\theta$ in the $(Z,X)$ plane. The polarization state is described by the Jones vector in the basis of linear $p$ and $s$ polarizations: $|\psi \rangle = \left( \begin{matrix} e_p \\ e_s \end{matrix} \right)$. The corresponding reflection coefficients have the form $R_{p,s}(\theta) = e^{i\alpha_{p,s}(\theta)}$ \cite{BornWolf}. We also introduce the local $(x,y,z)$ coordinate frames (with the common axis $y\equiv Y$) such that the $z$-axis follows the propagation directions of the incident and reflected beams, while the $x$ and $y$ axes correspond to the directions of $p$ and $s$ polarizations, respectively. The incident beam comprises a superposition of multiple plane waves with slightly different wavevectors ${\bf k}\simeq k \bar{\bf z} + k_x \bar{\bf x} + k_y \bar{\bf y}$, where $k_x^2 + k_y^2 \ll k^2$, and the overbar denotes the unit vectors of the corresponding axes. 
Taking into account slight deviations of the wavevectors from the central direction $k \bar{\bf z}$, and dependence of the reflection coefficient on the angle of incidence, the reflection can be described by a Jones-like matrix acting on the input Jones vector $|\psi_{\rm in} \rangle$ \cite{Bliokh2013JO}:
\begin{equation}
\label{S_beam}
\hat{S}\simeq
\!\left( \begin{matrix} e^{i\alpha_p}\!\left(1+i\kappa_x \frac{\partial \alpha_p}{\partial\theta} \right) &  \kappa_y\!\left(e^{i\alpha_p} + e^{i\alpha_s}\right)\!\cot\theta \\ -\kappa_y\!\left(e^{i\alpha_p} + e^{i\alpha_s}\right)\!\cot\theta & e^{i\alpha_s}\!\left(1+i\kappa_x \frac{\partial \alpha_s}{\partial\theta} \right) \end{matrix} \right)\!,
\end{equation}
where $\kappa_{x,y}=k_{x,y}/k$ are small parameters. 

%FFFFFFFFFFFFFFFFFFFFFFFFFFFFFFFFFFFFFFFFFFFFFF
%FFFFFFFFFFFFFFFFFFFFFFFFFFFFFFFFFFFFFFFFFFFFFF

The scattering matrix \eqref{S_beam} can be diagonalized by the transformation
\begin{equation}
\hat{U} \simeq \left( \begin{matrix} 1 &  i a \kappa_y\cot\theta \\ i a \kappa_y\cot\theta & 1 \end{matrix} \right)\!,~~
\hat{S}' = {\rm diag}\!\left( e^{i{\alpha}_1}, e^{i{\alpha_2}} \right)\!,
\label{U_beam}
\end{equation}
where $a=-i\,\dfrac{e^{i\alpha_s}+e^{i\alpha_p}}{e^{i\alpha_s}-e^{i\alpha_p}}= 
% \frac{\sin(\phi_p-\phi_s)}{1-\cos(\phi_p-\phi_s)}=
\cot({\tilde{\alpha}}/{2})$, $\tilde{\alpha}=\alpha_p - \alpha_s$, and $\alpha_{1,2} \simeq \alpha_{p,s} + \kappa_x \dfrac{\partial \alpha_{p,s}}{\partial\theta}$. 
Both $\hat{S}$ and $\hat{U}$ depend on the transverse wavevector components ${\bf k}_\perp = (k_x, k_y)$. Therefore, we introduce the corresponding GWSO $\hat{\bf Q}=i\hat{S}^\dag  \dfrac{\partial \hat{S}}{\partial {\bf k}_\perp}$, as well as the Berry connection 
\begin{equation}
\hat{\bf A} = i\hat{U}^\dag \dfrac{\partial \hat{U}}{\partial {\bf k}_\perp} \simeq -k^{-1}\! \left( \begin{matrix} 0 & 1 \\ 1 & 0 \end{matrix} \right) a \cot\theta\, \bar{\bf k}_y\,.
\label{A_beam}
\end{equation}

Substituting Eqs.~\eqref{U_beam} and \eqref{A_beam} into the general Eqs.~\eqref{Qd-Qg}, we derive the dynamic and geometric parts of the GWSO:
\begin{align}
\hat{\bf Q}_d &  
%\simeq \hat{\bf Q}'
\simeq - k^{-1}\! \left(\begin{matrix} \frac{\partial \alpha_{p}}{\partial\theta} & 0\\ 0 &\frac{\partial \alpha_{s}}{\partial\theta}\end{matrix}\right) \bar{\bf k}_x\,,\nonumber \\
\hat{\bf Q}_g & \simeq
k^{-1}\! \left( \begin{matrix} 0 & 1- e^{-i\tilde{\alpha}} \\ 1- e^{i\tilde{\alpha}} & 0 \end{matrix} \right)\! a \cot\theta\, \bar{\bf k}_y\,. 
\label{Qd-Qg_beam}
\end{align}
These parts are directed orthogonally to each other. Their expectation values with the incident wavefunction $|\psi_{\rm in}\rangle$ determine the orthogonal shifts in the transverse position of the reflected beam (see Fig.~\ref{Fig_beam}):
\begin{align}
\Delta {\bf r}_d & \simeq - k^{-1}\!\left[ |e_{p\,{\rm in}}|^2\frac{\partial \alpha_{p}}{\partial\theta} + |e_{s\,{\rm in}}|^2\frac{\partial \alpha_{s}}{\partial\theta}\right]\!\bar{\bf x}\,, \nonumber \\
\Delta {\bf r}_g & \simeq
k^{-1}\!\left[ \chi \sin\tilde{\alpha} - \sigma \left(1+\cos\tilde{\alpha}\right)\right]\cot\theta \,\bar{\bf y}\,, 
\label{Delta_R}
\end{align}
where $\chi = 2{\rm Re}(e_{p\,{\rm in}}^*e_{s\,{\rm in}})$ and $\sigma = 2{\rm Im}(e_{p\,{\rm in}}^*e_{s\,{\rm in}})$ are the Stokes parameters of the incident light, expressed now in the basis of linear polarizations.

The reflected-beam shifts \eqref{Delta_R} along the $x$ and $y$ axes are well known as the {\it Goos-H\"{a}nchen} and {\it Imbert-Fedorov (spin-Hall effect)} shifts, respectively \cite{Bliokh2013JO, Goos1947AP, Artmann1948AP, Bliokh2006PRL, Hosten2008S, Toppel2013NJP, Dennis2012NJP, Bliokh2016O}. \blue{Note that the Goos-H\"{a}nchen shift $\Delta{\bf r}_d$ obtained here has a sign opposite to that known in the literature. This discrepancy is instructive and stems from the use of the position operator $x = i\partial/\partial k_x$ in our formalism. However, reflection results in an inversion $k_x \to - k_x$ in the reflected beam \cite{Bliokh2013JO}, so we should have used the operator $x = -i\partial/\partial k_x$ for the shift of the reflected beam. Applying this sign correction to the first equations in Eqs.~\eqref{Qd-Qg_beam} and \eqref{Delta_R} restores the standard Goos-H\"{a}nchen shift, consistent with the literature.}

In summary, we have derived the reflected-beam shifts using the universal GWSO formalism and revealed their distinct physical origins. The Goos-H\"{a}nchen effect is of purely dynamic origin, arising from variations in the eigenvalues of the reflection matrix (via their phases $\alpha_{1,2}$).
In contrast, the Imbert–Fedorov shift has a purely geometric origin, associated with variations in the eigenvectors of the scattering matrix $\hat{S}$, and linked to geometric phases between the plane-wave components in the beam's spectrum \cite{Bliokh2013JO, Bliokh2015NP}.

%and Imbert-Fedorov effects have purely dynamic and geometric origins, respectively. Indeed, the Goos-H\"{a}nchen effect is associated with variations of the eigenvalues of the reflection matrix, via their phases $\alpha_{1,2}$, while the Imbert-Fedorov shift is related to variations of the eigenvectors of the $\hat{S}$-martix, and is underpinned by geometric phases between the plane waves in the beam spectrum \cite{Bliokh2013JO, Bliokh2015NP}. 
%Our formalism unveiled the dynamic and geometric origins of these effects via straightforward substitution of the scattering matrix into the general GWSO formalism \eqref{momentum_d_g} and \eqref{Qd-Qg}.

%%%%%%%%%%%%%%%%%%%%%%%%%%%%%%%%%%%%%%%%%%%%%%%%%%
\subsection{Wigner time delays in 1D scattering}
%%%%%%%%%%%%%%%%%%%%%%%%%%%%%%%%%%%%%%%%%%%%%%%%%%

\begin{figure}[t!]
\centering
\includegraphics[width=\linewidth]{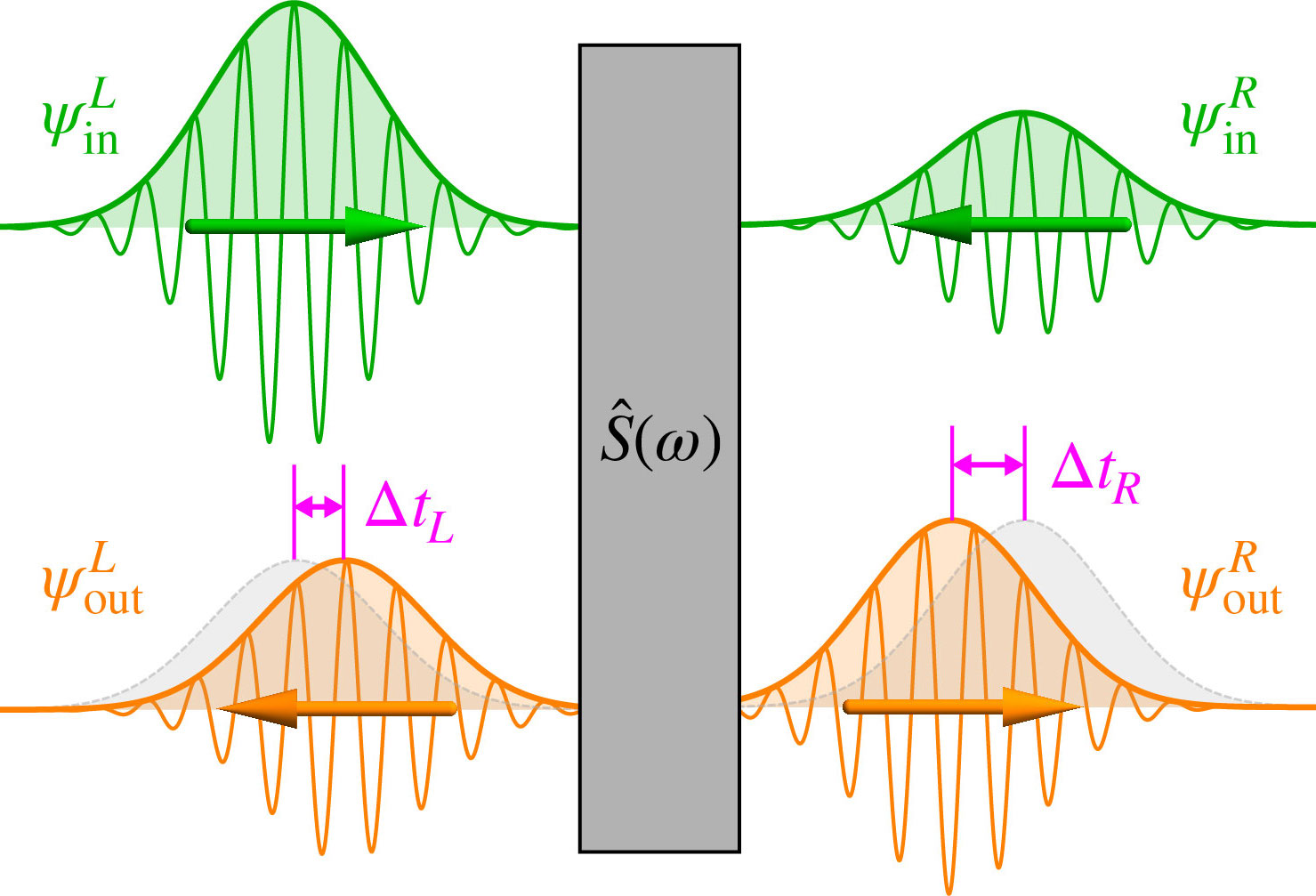}
\caption{Schematics of 1D wavepacket scattering described by a frequency-depended scattering matrix $\hat{S}(\omega)$, resulting in the Wigner time delays $\Delta t$ for the scattered wavepackets. The gray profiles show the expected output wavepackets assuming a non-dispersive scattering matrix $\hat{S}(\omega_c)$, evaluated at the central frequency $\omega=\omega_c$ of the incident wavepackets.} 
\label{Fig_resonator} 
\end{figure}

Finally, we consider the Wigner time delays (shifts), which originally motivated the introduction of the Wigner-Smith operator \cite{Wigner1955PR, Smith1960PR, Winful2006PR, Chiao1997PO}. 
A simple example is provided by the 1D scattering of a wavepacket (sufficiently narrow in frequency bandwidth, i.e., sufficiently extended in time) at a scatterer, such as a quantum-mechanical potential barrier or an optical resonator, see Fig.~\ref{Fig_resonator}. 
There is a pair of incoming waves (from two sides of the scatterer) and a pair of outgoing waves, so that the wavefunction can be written as $|\psi\rangle = \left( \begin{matrix} \psi^L \\ \psi^R \end{matrix} \right)$, where $\psi^L$ and $\psi^R$ are the waves on the left and right of the scatterer, respectively. The corresponding unitary scattering matrix (assuming time-reversal symmetry) has the form \cite{Tannor_book} 
\begin{equation}
\hat{S} = e^{i\varphi}\left( \begin{matrix} e^{i\delta} r & \sqrt{1-r^2}  \\ \sqrt{1-r^2} & -e^{-i\delta} r  \end{matrix} \right),
\label{S_1D}
\end{equation}
where $r\in [0,1]$ is an effective reflection coefficient, whereas $\varphi$ and $\delta$ are phase parameters. We assume that all parameters depend on the wave frequency $\omega$, so that the corresponding dispersion-induced time shifts of the output wavepackets are described by the GWSO $\hat{Q}=-i\hat{S}^\dag \dfrac{\partial \hat{S}}{\partial \omega}$. 

Note that the time shift in this context, $\Delta t$, represents appropriately averaged shifts of the intensity centroids of the two outgoing wavepackets, $\Delta t_{L,R}$. These shifts are defined with respect to the expected temporal positions of the outgoing wavepackets for a nondispersive scattering matrix $\hat{S}(\omega_c)$, where $\omega_c$ is the central frequency of the incoming wavepackets, see Fig.~\ref{Fig_resonator}.

%FFFFFFFFFFFFFFFFFFFFFFFFFFFFFFFFFFFFFFFFFFFFFF
%FFFFFFFFFFFFFFFFFFFFFFFFFFFFFFFFFFFFFFFFFFFFFF

If the scatterer is mirror-symmetric, then $S_{11}=S_{22}$, resulting in $\delta\equiv \pi/2$, and the matrix \eqref{S_1D} reduces to:
\begin{equation}
\hat{S}_{\rm sym} = e^{i\varphi}\left( \begin{matrix} ir & \sqrt{1-r^2}  \\ \sqrt{1-r^2} & ir  \end{matrix} \right).
\label{S_1D_s}
\end{equation}
This matrix can be diagonalized by a constant unitary transformation:
\begin{equation}
\hat{U} = \frac{1}{\sqrt{2}}\left(\begin{matrix} 1 & 1  \\ 1 & -1  \end{matrix} \right),\quad 
\hat{S}'={\rm diag}\!\left(e^{i(\varphi+\vartheta)},-e^{i(\varphi-\vartheta)}\right),
\label{U_1D_s}
\end{equation}
where $\vartheta = {\rm arcsin}(r)$. The eigenvectors of $\hat{S}_{\rm sym}$, which form the columns of $\hat{U}$, are constant. Therefore, all geometric contributions vanish: $\hat{Q}_g = \hat{A} = 0$. In this case, the dynamic GWSO coincides with the total GWSO and reads:
\begin{equation}
\hat{Q}=\hat{Q}_d = \frac{\partial \varphi}{\partial \omega} + \left(\begin{matrix} 0 & 1  \\ 1 & 0  \end{matrix} \right)
\frac{\partial \vartheta}{\partial \omega}\,.
\label{Qd_1D_s}
\end{equation}
The corresponding averaged time shift for the scattered wavepackets is given by the expectation value of $\hat{Q}$ with the incident wavefunction $|\psi_{\rm in} \rangle = \left( \begin{matrix} \psi^L_{\rm in} \\ \psi^R_{\rm in} \end{matrix}\right)$:
\begin{equation}
\Delta t = \Delta t_d= \frac{\partial \varphi}{\partial \omega} + \tau
\frac{\partial \vartheta}{\partial \omega}\,.
\label{Delta_t_s}
\end{equation}
Here we defined $\tau = 2\,{\rm Re} (\psi^{L*}_{\rm in} \psi^R_{\rm in})$, analogous to the Stokes parameter in polarization optics, and the frequency derivatives are evaluated at the central frequency $\omega=\omega_c$. For a wavepacket incident from one side only, $\tau=0$, the time shift reduces to the derivative of the total phase $\varphi$, which corresponds to the original Wigner formula.

Consider now a specific scatterer without mirror-symmetry, described by the scattering matrix \eqref{S_1D} with $\varphi \equiv 0$ and $\delta \equiv 0$: 
\begin{equation}
\hat{S}_{\rm asym} =\left( \begin{matrix} r & \sqrt{1-r^2}  \\ \sqrt{1-r^2} & - r  \end{matrix} \right).
\label{S_1D_a}
\end{equation}
We chose this matrix as a toy model that allows for a simple analytic solution. It has constant eigenvalues $s_{1,2}=\pm 1$, and therefore the dynamic contribution to the time shift vanishes: $\hat{Q}_d =0$. In turn, the eigenvectors of $\hat{S}_{\rm asym}$ depend on $r$, which results in the following diagonalization matrix:
\begin{equation}
\hat{U} = \frac{1}{\sqrt{2}}\left(\begin{matrix} \sqrt{1+r} & -\sqrt{1-r}  \\ \sqrt{1-r} & \sqrt{1+r}  \end{matrix} \right),~~
\hat{S}'={\rm diag}\!\left(1,-1\right).
\label{U_1D_a}
\end{equation}
The corresponding Berry connection is
\begin{equation}
\hat{A} =  -i \hat{U}^\dag \dfrac{\partial \hat{U}}{\partial \omega} = \frac{1}{2}\left( \begin{matrix} 0 & -i \\ i & 0\end{matrix} \right)\! \dfrac{\partial \vartheta}{\partial \omega}\,.
\label{A_1D_a}
\end{equation}
Substituting Eqs.~\eqref{U_1D_a} and \eqref{A_1D_a} into Eqs.~\eqref{Qd-Qg}, we obtain a purely geometric GWSO:
\begin{align}
\hat{Q} = \hat{Q}_g = \left( \begin{matrix} 0 & -i \\ i & 0\end{matrix} \right)\! \dfrac{\partial \vartheta}{\partial \omega}\,.
\label{Qg_1D_a}
\end{align}
This results in a purely geometric time shift:
\begin{equation}
\Delta t = \Delta t_g= \sigma
\frac{\partial \vartheta}{\partial \omega}\,,
\label{Delta_t_a}
\end{equation}
where $\sigma = 2\,{\rm Im} (\psi^{L*}_{\rm in} \psi^R_{\rm in})$. Thus, to observe the geometric time shift \eqref{Delta_t_a}, one has to consider incoming wavepackets from the left and right that are phase-shifted by $\pi/2$.

These examples demonstrate that even the simplest \red{scalar} 1D systems can exhibit both dynamic and geometric shifts of conjugate variables. \red{(This is due to the fact that the scattering problem has two channels and involves a two-level-like formalism.)}  Remarkably, the signs of dynamic and geometric time shifts \eqref{Delta_t_s} and \eqref{Delta_t_a}, proportional to the reflection-coefficient derivative $\partial\vartheta/\partial\omega$, can be controlled by the relative phase of wavepackets incident from the two sides: $\psi_R=\pm \psi_L$ ($\tau = \pm 1$) or $\psi_R=\pm i\psi_L$ ($\sigma = \pm 1$).

%%%%%%%%%%%%%%%%%%%%%%%%%%%%%%%%%%%%%%%%%%%%%%%%%%
\section{Conclusions}
%%%%%%%%%%%%%%%%%%%%%%%%%%%%%%%%%%%%%%%%%%%%%%%%%%

In this work, we have presented a unified theoretical framework that extends the geometric-dynamic decomposition (originally developed in the wave-evolution context) to general wave-scattering problems. Moreover, instead of analyzing the wave phase, we consider shifts in the expectation values of observables between the input and output wave states. Central to our analysis is the generalized Wigner-Smith operator (GWSO) involving gradients of the unitary scattering matrix with respect to physical parameters, such as time, position, frequency, or wavevector. This operator captures the expectation-values shifts of the conjugate variables: frequency, wavevector, time, or position.

Our key contribution is a gauge-invariant decomposition of these shifts into dynamic and geometric components: the former arising from gradients of the scattering matrix’s eigenvalues, and the latter from gradients of its eigenvectors, via the corresponding Berry connection.
%dynamic components, associated with gradients of the eigenvalues of the scattering matrix, and geometric components, linked to gradients of its eigenvectors via the corresponding Berry connection. 
We demonstrated the broad applicability of this decomposition through diverse examples, including time-varying waveplates, spatially varying metasurfaces, beam shifts at interfaces, and Wigner time delays in 1D scattering. 

Our approach not only recovers known phenomena (such as the Goos-Hänchen and Imbert-Fedorov shifts or the Pancharatnam-Berry phase) within a unified formalism, but also reveals new physical features,  provides new insights, and uncovers previously unnoticed interconnections. For instance, our geometric-dynamic decomposition of frequency shifts in polarized-light transmission through a time-varying waveplate provides a more fundamental description than the conventional Pancharatnam–Berry phase framework: it does not require cyclic evolution and remains gauge-invariant at every point of the base parameter space. 
Importantly, identifying the geometric and dynamic origins of various scattering-induced shifts offers new ways to control wave-scattering phenomena by tailoring the eigenvectors and eigenvalues of the scattering matrix.

%Notably, our analysis reveals that geometric effects can be dominant or exclusive in systems where eigenvalue variations vanish, and that the gauge-invariant separation we introduce offers conceptual and computational advantages over traditional phase-based methods.

Looking ahead, the presented approach offers several promising avenues for further exploration. One important direction is the extension of this formalism to non-Hermitian and open systems \cite{Moiseyev_book, El-Ganainy2018NP, Ozdemir2019NM}, where the scattering matrix becomes non-unitary and complex shifts, as well as singularities associated with exceptional points, may arise \cite{Asano2016NC, Giovannelli2024, byrnes2024perturbingscatteringresonancesnonhermitian}. 
Another important extension involves anomalously large shifts in the context of `quantum weak measurements' \cite{Kofman2012PR, Dressel2014RMP, Dennis2012NJP, Asano2016NC} where the output wave undergoes `post-selection', i.e., projection on a specifically chosen final state.   
Furthermore, integrating the GWSO-based decomposition into inverse design and machine-learning strategies could enable intelligent control of scattering in disordered or structured media \cite{Horodynski2022}, optimizing wavefronts for applications ranging from imaging \cite{Pai2021NP, Matthes2021PRX, Ber2025} to micromanipulation \cite{Hupfl2023PRL}.

In summary, the decomposition of scattering-induced shifts into geometric and dynamic components provides a powerful and unifying perspective through which a wide array of wave phenomena can be analyzed and engineered.

%Our approach provides a new unified view and important insights in all these problems, and it can be straightforwardly applied to numerous other wave-scattering systems. 

%%%%%%%%%%%%%%%%%%%%%%%%%%%%%%
\section*{Acknowledgements} 
%%%%%%%%%%%%%%%%%%%%%%%%%%%%%%
%\section*{Funding}
We acknowledge helpful discussions with Prof. Michael V. Berry and Prof. Miguel A. Alonso, as well as support from the Austrian Science Fund (FWF) [10.55776/COE1], Marie Sk\l{}odowska-Curie COFUND Programme of the European Commission (project HORIZON-MSCA-2022-COFUND-101126600-SmartBRAIN3), 
ENSEMBLE3 Project (MAB/2020/14) which is carried out within the International Research Agendas Programme (IRAP) of the Foundation for Polish Science co-financed by the European Union under the European Regional Development Fund and Teaming Horizon 2020 programme (GA. No. 857543) of the European Commission, and the project of the Minister of Science and Higher Education ``Support for the activities of Centers of Excellence established in Poland under the Horizon 2020 program'' (contract MEiN/2023/DIR/3797). 
%This research was funded in part by the Austrian Science Fund (FWF) [10.55776/COE1]. 
%This line should be added if we go for open access: For Open Access purposes, the authors have applied a CC BY public copyright license to any author accepted manuscript version arising from this submission.

\pagebreak
\newpage
%\onecolumngrid
\appendix

%%%%%%%%%%%%%%%%%%%%%%%%%%%%%%%%%%%%%%%%%%%%
\section{Time-varying waveplate and the Pancharatnam-Berry phase}
%%%%%%%%%%%%%%%%%%%%%%%%%%%%%%%%%%%%%%%%%%%%

Using the normalized Jones vector of the incident light, $|\psi_{\rm in}\rangle = \left( \begin{matrix} e^+_{\rm in} \\ e^-_{\rm in} \end{matrix} \right)$, and the scattering (Jones) matrix \eqref{S_plate}, the transmitted light is described by the wavefunction $|\psi_{\rm out}\rangle = \left( \begin{matrix} e^+_{\rm out} \\ e^-_{\rm out} \end{matrix} \right) = \hat{S} |\psi_{\rm out}\rangle $. The normalized Stokes parameters of the transmitted light are: $\tau'=2{\rm Re}(e^{+*}_{\rm out}e^-_{\rm out})$, $\chi'=2{\rm Im}(e^{+*}_{\rm out}e^-_{\rm out})$, and $\sigma'=|e^{+}_{\rm out}|^2-|e^-_{\rm out}|^2$; they can be expressed via the Stokes parameters of the incident light:
\begin{widetext}
\begin{align} 
\label{Stokes_transmitted}
%\tau' &=\left(\cos^2\!\frac{\delta}{2}+\sin^2\!\frac{\delta}{2}\cos 4\varphi \right)\tau + \sin^2\!\frac{\delta}{2}\sin 4\varphi\, \chi - \sin\delta \sin 2\varphi\, \sigma\,,
\tau' &=\left(\cos^2\!2\varphi+\cos\delta \sin^2\! 2\varphi \right)\tau + (1-\cos\delta)\sin 2\varphi \cos 2\varphi\, \chi - \sin\delta \sin 2\varphi\, \sigma\,, \nonumber \\
%
%\chi' &= \sin^2\!\frac{\delta}{2}\sin 4\varphi\, \tau + \left(\cos^2\!\frac{\delta}{2}-\sin^2\!\frac{\delta}{2}\cos 4\varphi \right)\chi + \sin\delta \cos 2\varphi\, \sigma\,, 
\chi' &= (1-\cos\delta)\sin 2\varphi \cos 2\varphi\, \tau
+ \left(\cos\delta\cos^2\!2\varphi+ \sin^2\! 2\varphi \right)\chi 
+ \sin\delta \cos 2\varphi\, \sigma\,, \\
\sigma' &= \sin\delta \sin 2\varphi\, \tau
- \sin\delta \cos 2\varphi\, \chi 
+ \cos \delta\, \sigma\,. \nonumber
\end{align}
\end{widetext}
These equations determine the Mueller matrix of the waveplate, which transforms the incident Stokes vector $\vec{\Sigma} = \left( \begin{matrix} \tau \\ \chi \\ \sigma \end{matrix} \right)$ into the transmitted Stokes vector $\vec{\Sigma}' = \left( \begin{matrix} \tau' \\ \chi' \\ \sigma' \end{matrix} \right)$. The unit Stokes vectors represent the polarization state of light on the Poincar\'{e} sphere \cite{Azzam_book}, Fig.~\ref{Fig_sphere}. Substituting the last equation \eqref{Stokes_transmitted} into Eqs.~\eqref{Delta_omega} of the main text, we obtain Eq.~\eqref{Coriolis} for the geometric frequency shift as the Coriolis or angular-Doppler effect. 

Let us consider an infinitesimal evolution of the waveplate between instants of time $t$ and $t+dt$, which drives the transmitted Stokes vector from $\vec{\Sigma}'$ to $\vec{\Sigma}'+d\vec{\Sigma}'$. The three vectors $\vec{\Sigma}$, $\vec{\Sigma}'$, and $\vec{\Sigma}'+d\vec{\Sigma}'$, together with the geodesic arcs connecting them, form a triangle on the Poincar\'{e} sphere, Fig.~\ref{Fig_sphere}. The Pancharatnam-Berry (PB) phase corresponding to this infinitesimal evolution is equal to minus half of the solid angle cut by this triangle. It can be written as \cite{VanOosterom1983, Eriksson1990, BAD2019}:
\begin{equation}
\label{PB_def}
d\Phi_{PB} = - \frac{(\vec{\Sigma}\times \vec{\Sigma}')\cdot d\vec{\Sigma}'}{2(1+\vec{\Sigma}\cdot \vec{\Sigma}')}\,.
\end{equation} 
One can use $d\vec{\Sigma}' = \dfrac{\partial \vec{\Sigma}'}{\partial \delta} d\delta + \dfrac{\partial \vec{\Sigma}'}{\partial \varphi} d\varphi$, which indicates that the PB phase is generally related to variations of both the `dynamic' parameter $\delta$ and `geometric' parameter $\varphi$.
Finding the general relation between the PB phase and frequency shifts, Eqs.~\eqref{Delta_omega} in the main text, is challenging; therefore we consider specific simplified cases. 

First, for a circularly-polarized incident light, $\vec{\Sigma} = \left( \begin{matrix} 0 \\ 0 \\ \sigma \end{matrix} \right)$,
calculations of Eqs.~\eqref{PB_def} and \eqref{Stokes_transmitted} yield:
\begin{align}
\label{PB_sigma}
d\Phi_{PB} = \sigma(1-\cos\delta)\,d\varphi =
- \Delta\omega_g\,dt\,.
\end{align} 
In this case, the PB phase \eqref{PB_sigma} is determined by the geometric frequency shift related to variations of the parameter $\varphi(t)$, whereas the dynamic contribution vanishes: $\Delta\omega_d=0$. See Fig.~\ref{Fig_sphere}(a) in the main text.

Second, for a vertically or horizontally polarized incident light, $\vec{\Sigma} = \left( \begin{matrix} \tau \\ 0 \\ 0 \end{matrix} \right)$, we obtain:
\begin{align}
\label{PB_tau}
d\Phi_{PB} & =  - D_\tau\,\tau\left( -\frac{1}{2}\cos 2\varphi \, d\delta + \sin 2\varphi \sin\delta \, d\varphi\right) \nonumber\\
& = - D_\tau (\Delta\omega_d+\Delta\omega_g)\,dt\,,
\end{align} 
where $D_\tau =\dfrac{1-\tau'\tau}{1+\tau'\tau}$. In this case, the PB phase \eqref{PB_tau} is proportional to the sum of the geometric and dynamic frequency shifts \eqref{Delta_omega}, i.e., to the total frequency shift. 

Finally, for a diagonally-polarized incident light, $\vec{\Sigma} = \left( \begin{matrix} 0 \\ \chi \\ 0 \end{matrix} \right)$, we derive:
\begin{align}
\label{PB_chi}
d\Phi_{PB} & =  - D_\chi\,\chi\left( -\frac{1}{2}\sin 2\varphi \, d\delta + \cos 2\varphi \sin\delta \, d\varphi\right) \nonumber\\
& =- D_\chi (\Delta\omega_d+\Delta\omega_g)\,dt\,,
\end{align} 
where $D_\chi =\dfrac{1-\chi'\chi}{1+\chi'\chi}$. The PB phase \eqref{PB_chi} is also proportional to the sum of the geometric and dynamic frequency shifts \eqref{Delta_omega}. 

Thus, Eqs.~\eqref{PB_sigma}--\eqref{PB_chi} show that the PB phase is related to the {\it total} frequency shift induced by variations of the parameters $\delta(t)$ and $\varphi(t)$. This phase is `geometric' only in the sense of its relation to the solid angle on the Poincar\'{e} sphere. %Our considerations show that in fact it consists of the dynamical and geometric contributions related to variations of the eigenvalues and eigenvectors of the scattering (Jones) matrix of the wave plate. 
Moreover, the coefficients $D_\tau$ and $D_\chi$ in Eqs.~\eqref{PB_tau} and \eqref{PB_chi} show that there is no universal relation between the observable frequency shift and the PB phase (i.e., solid angles on the Poincar\'{e} sphere). 
%This agrees with the conclusions of \cite{Malykin2004PU}.

For the incident $\tau$-linear or $\chi$-linear polarizations, $1\pm\vec{\Sigma}\cdot \vec{\Sigma}' = 1\pm\tau'\tau$ or $1\pm\vec{\Sigma}\cdot \vec{\Sigma}' = 1\pm\chi'\chi$. Therefore, in these cases one can remove the $D_\tau$ and $D_\chi$ factors by using an alternative definition of the PB phase:
\begin{equation}
\label{PB_def_2}
d\tilde{\Phi}_{PB} = - \frac{(\vec{\Sigma}\times \vec{\Sigma}')\cdot d\vec{\Sigma}'}{2(1-\vec{\Sigma}\cdot \vec{\Sigma}')}\,,
\end{equation} 
which corresponds to plus half of the solid angle formed by the Stokes vectors $-\vec{\Sigma}$, $\vec{\Sigma}'$, and $\vec{\Sigma}'+d\vec{\Sigma}'$ \cite{BAD2019}, Fig.~3(b)
For this definition, Eqs.~\eqref{PB_tau} and \eqref{PB_chi} take the form:
\begin{align}
\label{PB_tau_2}
d\tilde{\Phi}_{PB} = - (\Delta\omega_d+\Delta\omega_g)\,dt\,.
\end{align} 
However, in this case Eq.~\eqref{PB_sigma} becomes
\begin{align}
\label{PB_sigma_2}
d\tilde{\Phi}_{PB} =
- \tilde{D}_\sigma \Delta\omega_g\,dt\,,
\end{align} 
where $\tilde{D}_\sigma =\dfrac{1+\sigma' \sigma}{1-\sigma' \sigma}$.

%%%%%%%%%%%%%%%%%%%%%%%%%%%%%%%%%%%%%%%%%%%%
\section{Another spherical representation for time-varying waveplate}
%%%%%%%%%%%%%%%%%%%%%%%%%%%%%%%%%%%%%%%%%%%%

Since the total frequency shift $\Delta\omega$ consists of the two contributions proportional to $\partial\delta/\partial t$ and $\partial\varphi/\partial t$, Eqs.~\eqref{Delta_omega}, the phase shift accumulated during evolution of the waveplate can be written in the form $\Delta\Phi = - \int_0^{t} \Delta\omega\, dt = \int ({\mathcal A}_\delta d\delta + {\mathcal A}_\varphi d\varphi)$, i.e., as a contour integral of the corresponding vector field $({\mathcal A}_\delta,{\mathcal A}_\varphi)$ in the $(\delta,\varphi)$ space. This representation acquires an interesting form if we assume that this space is a \textit{sphere} with spherical angles $(\theta,\phi)=(\delta,2\varphi)$. In this case, we obtain: %$\Delta\Phi = \int ({\mathcal A}_\theta\, d\theta + {\mathcal A}_\phi \sin\theta\, d\phi)$:
\begin{align}
\label{A_sphere}
\Delta\Phi  & = \int ({\mathcal A}_\theta\, d\theta + {\mathcal A}_\phi \sin\theta\, d\phi)\,,\nonumber\\
{\mathcal A}_\theta  & = \frac{1}{2}(\tau\cos\phi+\chi\sin\phi)\,, \nonumber \\
{\mathcal A}_\phi & = \frac{1}{2}\!\left(\sigma\frac{1-\cos\theta}{\sin\theta}-\tau\sin\phi+\chi\cos\phi \right).
\end{align} 
Here the $\sigma$-dependent term in ${\mathcal A}_\phi$ corresponds to half of the standard connection on the sphere. 

For a cyclic evolution on the $(\theta,\phi)$-sphere, the contour integral can be transformed into the corresponding surface integral: 
%$\Delta\Phi = \iint {\mathcal F}  \sin\theta\,d\theta\, d\phi$
%
\begin{align}
\label{F_sphere}
\Delta\Phi & = \iint {\mathcal F}  \sin\theta\,d\theta\, d\phi\,,\nonumber\\
{\mathcal F} & = \frac{1}{\sin\theta}\frac{\partial}{\partial\theta}({\mathcal A}_\phi \sin\theta) -
\frac{1}{\sin\theta}\frac{\partial {\mathcal A}_\theta}{\partial\phi} \nonumber\\ 
& = \frac{1}{2}\!\left(\sigma + \tau\frac{\sin\phi}{\sin\theta}-\chi\frac{\cos\phi}{\sin\theta} \right).
\end{align} 
Here the $\sigma$-term corresponds to half of the solid angle enclosed by the contour on the $(\theta,\phi)$-sphere. Thus, for a circularly-polarized incident light with $\tau=\chi=0$, the total phase acquires a form similar to the PB phase, but now on the $(\theta,\phi)$-sphere instead of the Poincar\'{e} sphere. For a non-circular incident polarization, there is no such simple geometric interpretation. This agrees with the main conclusions from the consideration of the PB phase on the Poincar\'{e} sphere.  

Note the difference between the two representations. The Poincar\'{e} sphere represents the polarization $(\tau,\chi,\sigma)$-states of the incident and transmitted light, which depend on the waveplate parameters $(\delta,\varphi)$. Calculations of the PB phase on this sphere use a fixed standard connection and curvature, corresponding to the solid angle. In contrast, the $(\theta,\phi)$-sphere represents the $(\delta,\varphi)$-state of the plate. Calculations of the observable phase $\Delta\Phi$ on this sphere involve an effective connection  $({\mathcal{A}}_\theta, {\mathcal{A}}_\phi)$ and curvature ${\mathcal{F}}$ depending on the polarization $(\tau,\chi,\sigma)$-state of the incident light. In both representations, only the case of circular incoming polarization allows clear geometric interpretations.

\bibliography{refs}

\end{document}